\documentclass[reprint, amsmath, amssymb, aps, pre]{revtex4-2}
\usepackage{graphicx, dcolumn, bm, amsmath, amssymb, xcolor, enumitem, float}
\usepackage[spaces]{grffile}
\allowdisplaybreaks
\usepackage{bm,bbm}

\let\tempvec\vec
\renewcommand{\vec}[1]{\tempvec{{}#1}}
\let\tempbar\bar
\renewcommand{\bar}[1]{\tempbar{{}#1}}

\newcommand{\rv}{{\bf r}}
\newcommand{\xv}{{\bf x}}
\newcommand{\yv}{{\bf y}}
\newcommand{\nv}{\hat{\bf n}}
\newcommand{\Nv}{\hat{\bf N}}
\newcommand{\Tv}{\hat{\bf T}}
\newcommand{\ev}{\hat{\bf e}}

\newcommand{\uv}{{\bf u}}
\newcommand{\kapv}{\bm{\kappa}}
\newcommand{\bev}{\bm{\beta}}

\newcommand{\MW}[1]{{\color{red}{#1}}}

\begin{document}
    \title{Self-limiting states of polar misfits: Frustrated assembly of warped-jigsaw particles}
	\author{Michael Wang$^{1,2}$}
		\email{mwang@mail.pse.umass.edu}
	\author{Gregory Grason$^{2}$}
        \email{grason@umass.edu}
    \affiliation{$^{1}$Aix Marseille Univ, Université de Toulon, CNRS, CPT (UMR 7332), Turing Center for Living Systems, Marseille, France}
	\affiliation{$^{2}$Department of Polymer Science and Engineering, University of Massachusetts, Amherst, MA 01003}
	
	\begin{abstract}
		We study the ground state thermodynamics of a model class of geometrically frustrated assemblies, known as {\it warped-jigsaw} particles.  While it is known that frustration in soft matter assemblies has the ability to propagate up to mesoscopic, multi-particle size scales, notably through the selection of self-limiting domain, little is understood about how the symmetry of shape-misfit at the particle scale influences emergent morphologies at the mesoscale.  Here we show that polarity in the shape-misfit of warped-jigsaw puzzles manifests at a larger scale in the morphology and thermodynamics of the ground-state assembly of self-limiting domains.  We use a combination of continuum theory and discrete particle simulations to show that the polar misfit gives rise to two mesoscopically distinct polar, self-limiting ribbon domains.  Thermodynamic selection between the two ribbon morphologies is controlled by a combination of the binding anisotropy along distinct neighbor directions and the orientation of polar shape-misfit.  These predictions are valuable as design features for ongoing efforts to program self-limiting assemblies through the synthesis of intentionally frustrated particles, and further suggests a generic classification of frustrated assembly behavior in terms of the relative symmetries of shape-misfit and the underlying long-range inter-particle order it frustrates. 
	\end{abstract}
 
	\maketitle
 
    \section{\label{sec:introduction}Introduction}
        Geometrically frustrated assembly (GFA) has recently emerged as an organizing principle for complex, self-organized structures in soft matter.  In geometrically frustrated systems, such as in the frustrated ordering of magnetic spins \cite{Wannier:1950,Collins:1997,Vannimenus:1977}, a locally preferred arrangement of constituents is not compatible with the constraints on uniformly filling space.  In soft matter assemblies, this takes the form of molecular or particulate subunits whose shape and interactions favor a local ``misfit" with the longer-range inter-particle order.  This basic paradigm applies to a broad range of soft matter systems, including chiral and polar phases of liquid crystals \cite{Reddy:2006,Takezoe:2006,Fernandez-Rico:2020}, chiral membranes \cite{Ghafouri:2005,Armon:2014,Zhang:2019,Matsumoto:2009,Gibaud:2012,Sharma:2014} and protein bundles \cite{Aggeli:2001,Turner:2003,Grason:2007,Yang:2010,Brown:2014,Hall:2016}, colloidal crystals on curved surfaces \cite{Meng:2014,Irvine:2010,Guerra:2018,Li:2019}, and ill-fitting colloidal and nanoparticle assemblies \cite{Lenz:2017,Leroy:2023,Berengut:2020,Uchida:2017,Tyukodi:2022,Tanjeem:2022,Spivack:2022,Hall:2023,Serafin:2021}.  The key effect of frustration is to introduce {\it intra-assembly} gradients of misfit that extend to multi-subunit scales, leading to elastic costs that {\it accumulate} with increasing domain sizes.  This accumulation of elastic energy \cite{Meiri:2021} with domain size notably results in self-limitation, in which the equilibrium assembly states have well-defined, finite, multi-subunit sizes, a behavior that is not possible for nearly all other classes of equilibrium assembly \cite{Hagan:2021,Grason:2016}.

        These exotic and potentially useful behaviors of GFAs, in combination with an increasing array of synthetic techniques for fabricating geometrically-programmable assembly units \cite{Sigl:2021,Hayakawa:2022,Suzuki:2016,Berengut:2020,Uchida:2017}, have inspired theoretical efforts that attempt to connect microscopic features of the subunits and their interactions to the emergent larger-scale morphologies \cite{Hall:2023,Tyukodi:2022,Spivack:2022,Sullivan:2024,Wang:2024}.  Such studies, which range from continuum theories to numerical simulations of discrete misfitting subunit models, address basic and still broadly open questions: Given a ``misfitting'' subunit design, under what conditions does self-limited assembly occur, and what morphologies emerge from it?  From the point of view of engineering self-limiting states, it is particularly important to understand what design features of ``misfit'', such as particle shape and interactions, control assembly thermodynamics at the largest possible scales and ultimately, to translate those features to synthetically accessible modifications of subunits.

    	\begin{figure*}
    		\centering
    		\includegraphics[width=	\textwidth]{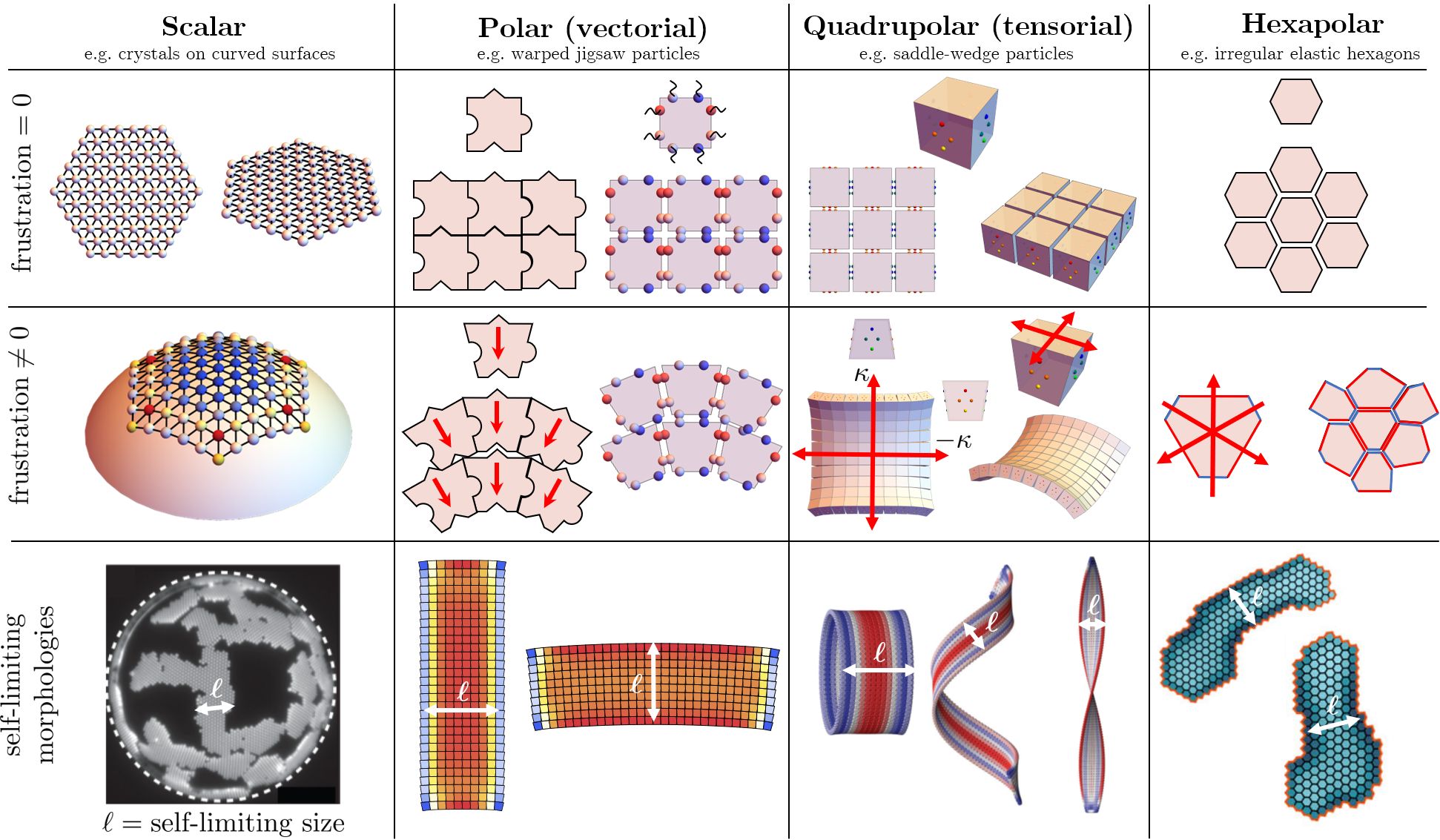}
    		\caption{Classification of geometrically frustrated assemblies in terms the symmetry or directionality of the frustration due to shape distortion and its mismatch with the underlying frustration free inter-particle order.  Top row shows unfrustrated assemblies and their preferred arrangement while the middle row shows corresponding frustrated assemblies and their distortions (position, orientation, and shape) from those preferred arrangements.  The bottom row shows example realizations of self-limiting assemblies.  Scalar frustration is represented by the crystallization on spherical surfaces as in Ref.\ \cite{Meng:2014}.  Vectorial frustration can be represented by assemblies of warped particles with a directed shape asymmetry.  Quadrupolar frustration is represented by double wedge particles that prefer locally nonzero Guassian curvature from Ref.\ \cite{Hall:2023}.  Hexapolar frustration is represented by warped hexagonal particles with long and short edges from Ref.\ \cite{Leroy:2023}.}
    		\label{fig:intro}
    	\end{figure*}

        It is in this context that we consider a basic classification of distinct GFA mechanisms in terms of their {\it misfit symmetry}, shown schematically in Fig.\ \ref{fig:intro} for the case of two-dimensional assemblies \footnote{This classification in terms of the shape-misfit symmetry may very well be extended beyond 2D}.  This notion applies to the cases where the degree of frustration may be continuously increased, through some experimental parameter, from the frustration-free cases where there is some type of long-range order such as in 2D hexagonal or square crystalline packing.  The introduction of frustration takes the form a local shape distortion incompatible with the (unfrustrated) background order.  The symmetry of this local distortion, and in particular its mismatch with the underlying frustration-free order, is what we refer to as the misfit symmetry.  For example, for the case of crystalline assemblies on spherical surfaces \cite{Meng:2014,Irvine:2010,Guerra:2018,Li:2019}, the underlying (frustration-free) order corresponds to a hexagonal lattice which uniformly tiles the 2D plane.  Imposing positive Gaussian curvature on the assembly, say by forcing particles to assemble on a spherical drop, disrupts the uniform hexagonal lattice order, since equilateral triangles do not (in general) tile spherical surfaces.  Since there is no particular directionality of this distortion of hexagonal order due the spherical curvature, we may refer to this as {\it scalar} misfit.  This is distinct from the other examples of directional misfit shown in Fig.\ \ref{fig:intro}, in which the misfit is characterized by distortions along one or more specific axes in the 2D subspace of the assembly.  For example, in the model of irregular hexagon assembly developed \cite{Lenz:2017,Leroy:2023}, frustration is introduced by shortening and lengthening alternate edges of an otherwise hexagonal elastic tile.  We can characterize the deformation of an unfrustated 6-fold symmetric particle into a lower 3-fold symmetry particle as {\it hexapolar} misfit.  Likewise, we may consider the hyperbolic membranes of ``saddle-wedge'' particles studied in Ref.\ \cite{Hall:2023}, whose shape and interactions favor a 2D square lattice assembly that is frustrated by a simultaneous preference for opposite curvatures in two orthogonal directions.  Hence, the preference for negatively-curved, minimal surface shapes of an otherwise crystalline membrane can be characterized as a quadrupolar or tensorial misfit.

        A central question, which we address in this article for a specific model of {\it polar} frustration in so-called {\it warped-jigsaw} particles \cite{Spivack:2022}, is how does the misfit symmetry of the building blocks impact the morphologies of self-limiting states that are selected by their assembly thermodynamics?  Previous theoretical and computational studies of a diverse range of GFA models found that 2D minimal energy morphologies of self-limiting states tend to break the symmetry of the underlying background order.  For example, while low energy shapes of hexagonal crystallites favor generally compact and six-fold symmetric shapes that optimize edge energies, the growth of elastic energy with domain size of crystals on a sphere favors strip-like morphologies \cite{Schneider:2005,Meng:2014} that grow arbitrarily long in one direction but remain narrow in the other (see example morphologies in Fig.\ \ref{fig:intro}).  The basic mechanism that promotes domain anisotropy, sometimes referred to as ``filamentation'', stems from the elasticity of frustrated domains, which requires inter-domain gradients in the elastic stress.  From an energetic point of view, it is favorable to orient the strain gradients along the narrow direction of domain, so that the range of strains (i.e.\ difference between maximal and minimal strains) remains small.  Hence, such scalar mechanisms of GFA, like crystalline assemblies on spherical surfaces, favor strip-like morphologies that break the underlying six-fold symmetry of the unfrustrated assembly, leading to multiple degenerate states (e.g.\ degenerate strip morphologies associated with the 3 low-energy growth directions).  A similar phenomenon occurs in frustrated hyperbolic crystalline \cite{Hall:2023} and irregular elastic hexagon assembly \cite{Lenz:2017,Leroy:2023}, in which optimal self-limiting morphologies break symmetry by remaining narrow in one direction of their 2D assembly and growing long in the other.  Notably, for these examples (e.g.\ scalar, quadrupolar, hexapolar), the misfit symmetries are commensurate with broken symmetries of self-limiting morphologies.  Hence, for these symmetries the possible resulting ``ribbon-like'' morphologies remain degenerate, and the physical assembly processes must spontaneously break symmetry to form self-limiting states.

        In this article, we investigate the geometrically-frustrated assembly and self-limiting morphologies of a polar misfit model of warped jigsaw (WJ) particles, and show that the polarity of local misfit profoundly reshapes the energetic landscape of optimal morphologies.  Specifically, we show that, unlike the scalar, quadrupolar or hexapolar cases, the effect of misfit polarity on the elastic energy landscapes generically lifts the degeneracy between multiple competing strip-like morphologies.  In general, we find that these non-degenerate morphologies are distinguished by the type of mechanical strain-gradients they develop along their narrow dimensions which is controlled by the relative angle between the polarity of WJ shape and their narrow dimension. As we will see, the strain-gradient patterns are characterized by an competition between intra-row bending and either inter-row shears or intra-row stretching.  
        
        In this article, we first consider the case where polar misfit is aligned with a lattice direction and find that the distinguishable mesoscopic morphologies have distinct patterns of local gradients, reflected in the polarity of the large scale assembly of either straight but polar strips or finite thickness, circular strips, as shown in Fig.\ \ref{fig:intro}.  Additionally, these distinct deformation patterns are characterized by distinct thickness dependencies of the elastic costs of frustration.  While both frustrated ribbon morphologies smoothly approach a uniformly strained bulk state at large thicknesses, they approach this ``frustration escape'' limit via two distinct thermodynamic routes, with accumulating costs of frustrations screened exponentially beyond a finite boundary for straight-polar ribbons, while the frustration costs in circular strips exhibit a power-law approach to the uniform bulk state.  We study how this anisotropy in elastic costs of frustrations competes with anisotropy in edge binding, and predict a phase diagram of ground-state morphologies that determines the stable regions of both straight-polar and circular finite-thickness ribbons, as well as a uniformly strained bulk (unlimited) state.  
        
        We next consider the case when misfit polarity is misaligned with the underlying two-dimensional lattice, which we show leads to a complex evolution of finite-width strip morphologies that mix features of the straight-polar and circular ribbon states.  Last, we show that while the elastic cost of frustration is generically smallest for narrow dimensions that are orthogonal to misfit polarity (i.e.\ straight-polar ribbons), ground-state ribbons generically favor growth along lattice directions due orientation dependence of the cohesive edge cost.  

        The article is organized as follows.  In Sec.\ \ref{sec:model}, we introduce a microscopic model of WJ particle assembly and the ingredients of a discrete subunit computational model of the elastic ground-states.  We then describe the pairwise inter-particle deformation modes and their connection to a continuum elastic formulation of the elastic energy of assembled domains of WJ particles.  Then in Sec.\ \ref{sec:psizero}, we study the case of misfit polarity aligned to one the 2D binding directions, and analyze the elastic and thermodynamic landscape of finite domains of WJ particles, showing the emergence of two distinct, self-limiting morphologies, characterized by distinct patterns of strain gradients.  Using this analysis of the energetics, we construct a phase diagram of the ground-state morphologies in terms of the relative edge binding strengths.  In Sec.\ \ref{sec:psinonzero}, we analyze and compare examples of misfit polarity which are not aligned to either lattice direction.  We first consider domains whose free edges are aligned along binding directions before generalizing to free edges that align with off-lattice directions to show that thermodynamically selected ribbons generically favor extending along the WJ particle binding directions.  We conclude with a discussion of the implications on the broader emergent effects of frustration on assembly, as well as for the engineering of self-limiting assemblies through design and fabrication of intentionally misfitting subunits.

    \section{\label{sec:model}Warped-jigsaw particle model}
        \begin{figure*}
            \centering
            \includegraphics[width=\textwidth]{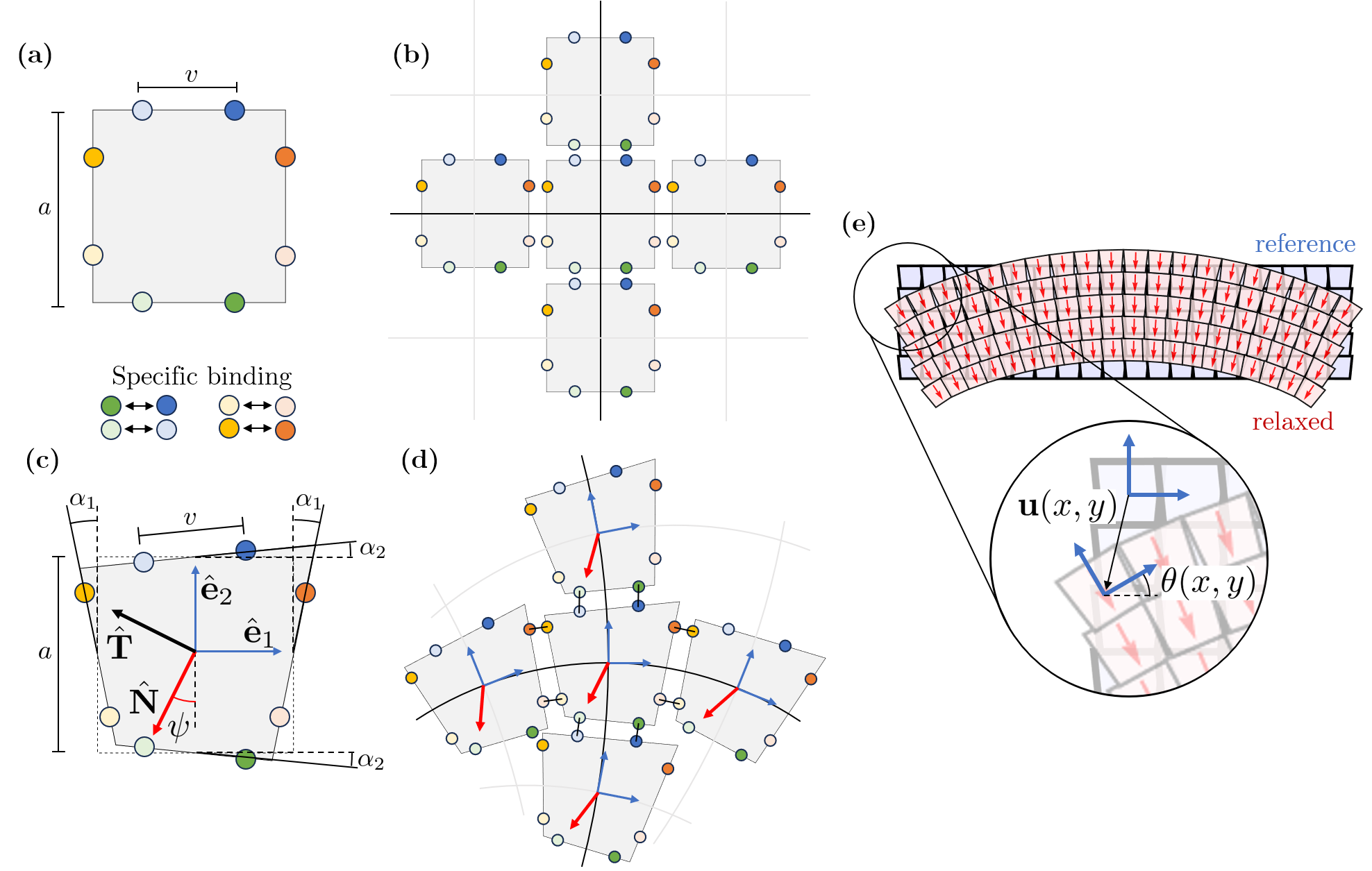}
            \caption{Discrete subunit model for warped-jigsaw particles.  \textbf{a)} Unfrustrated square subunit.  Colored circles indicate elastic binding sites with specific binding between dark green and dark blue, light green and light blue, dark yellow and dark orange, and finally light yellow and light orange.  The pairs of binding sites on each edge are separated by a distance $v$ while the overall subunit size is $a$.  \textbf{b)} Unfrustrated square subunits bind to form a square lattice.  \textbf{c)} and \textbf{d)} Frustrated quadrilateral subunit and the preferred binding horizontal and vertical binding of neighbors.  The unit vector $\Nv$ (red arrows) characterizes the direction of preferred curvatures $\boldsymbol{\kappa}_0$ i.e.\ the rotation of $\ev_1$ and $\ev_2$ along directions perpendicular to $\Nv$, described by the unit vector $\Tv$.  \textbf{e)}  Deformation of an assembly (light red) away from its uniform 2D square lattice reference state (light blue), resulting in a displacement field $\uv(x,y)$ and rotation field $\theta(x,y)$.}
            \label{fig:model}
        \end{figure*}
        Here, we introduce the warped-jigsaw (WJ) particle model of Spivack and coworkers \cite{Spivack:2022}.  The shape of WJ particles is described by rigid quadrilaterals in the 2D plane, shown in Fig.\ \ref{fig:model}.  The geometry and interactions of the WJ particles simultaneously favor binding along rows and columns in a 2D square lattice as well as gradients in particle orientation along a certain in-plane direction, which frustrates the uniform lattice order.  Each edge has a pair of elastic binding sites with specific interactions that bind together particles along two orthogonal directions: ``left-right'' (dark red-dark orange and light red-light orange sites) and ``top-bottom'' (dark blue-dark green and light blue-light green sites) edges of neighboring particles.  Such specific interactions, of the type that can be generated e.g.\ via sequence-specific DNA linker binding \cite{Cui:2022,Rogers:2011}, imply that the left (bottom) edges only bind favorably to the right (top) ones {\it and} that bound neighbors favor a certain relative orientation.  The relative orientations are described in a local frame of orthonormal unit directors $\{ \hat{{\bf e}}_1, \hat{{\bf e}}_2 \}$ aligned to the respective horizontal and vertical directions of a particle, but can rotate in the $xy$-plane within an assembly.  In the unfrustrated state, the particle is simply a square (Fig.\ \ref{fig:model}a) and when bound to neighbors, the ground state arrangement is that of a square lattice (Fig.\ \ref{fig:model}b), with the frames of all particles aligned uniformly.  
        
        Frustration is introduced by shape asymmetry, as shown in Fig.\ \ref{fig:model}c, where opposite edges of the unfrustrated square shape are skewed away from each other, forming taper angles $\alpha_1$ and $\alpha_2$ along the horizontal $\ev_1$ and vertical $\ev_2$ directions, respectively.  This shape asymmetry due to the taper angles results in a preferred relative rotation between neighboring particles and consequently, nonzero preferred curvatures along columns and rows of these particles (Fig.\ \ref{fig:model}d), a local arrangement that is incompatible with the positional order (square lattice) of the unfrustrated state.  As outlined in the introduction, this frustration is polar or \textit{vectorial} in nature in that it has a direction and magnitude.  We characterize the polarity of shape-misfit in terms of the {\it preferred curvature vector}
        \begin{equation}
            \kapv_0 = \kappa_0\Nv,
        \end{equation}
        where $\kappa_0$ is magnitude of preferred curvature and $\Nv$ characterizes the preferred bending direction of rows or columns of particles.  The resulting direction of rotation of the particle frames along a row or column can be described by the unit vector $\Tv=-\hat{\mathbf{z}}\times\Nv$, where $\hat{\mathbf{z}}$ is normal to $xy$ plane of assembly.   
        The preferred curvatures along rows and columns are given by $\kappa_i \equiv \kapv_0 \cdot \hat{\mathbf{e}}_i=(2\tan\alpha_i)/a$, which describe the amount of preferred bending orthogonal to local $\ev_i$ direction.  Hence, $\kappa_0 = \sqrt{ \kappa_1^2+\kappa_2^2}$ quantifies the {\it magnitude} of the misfit while $\Nv=-\sin\psi\,\hat{\mathbf{e}}_1-\cos\psi\,\hat{\mathbf{e}}_2$ describes its \textit{orientation}, which we parameterize by the angle $\psi$ in the local particle frame.   
        
        Shown in Fig.\ \ref{fig:model}e is an example of an assembly formed by trapezoidal WJ particles with $\psi = 0 $, whose shapes are symmetric along their $\hat{\mathbf{e}}_1$ (horizontal) directions and asymmetric along their $\hat{\mathbf{e}}_2$ (vertical) directions, as characterized by the downward misfit polarity $\Nv=-\hat{\mathbf{e}}_2$.  This asymmetry favors curvature along only the ``horizontal'' rows causing the assembly to deviate from a square lattice reference (unfrustrated) state.  To describe and model the elastic energy WJ assemblies, we take two approaches.  In the first, we perform energy minimizations of domains of elastically bound rigid WJ particles.  As we show, emergent ground states take the form of highly anisotropic rectangular domains with either finite numbers of columns $M$ or rows $N$.  We first assume that the domain boundaries align with binding directions so that domain edges are characterized by broken bonds along either the $\ev_1 $ or $\ev_2$ directions, although we relax this assumption and consider the possibility of domain edges aligning along different directions in Sec.\ \ref{sec:psinonzero}.  Cohesive costs of an exposed (unbound) WJ particle edge in direction ${\bf e}_\alpha$ is parameterized by the binding energy $\Sigma_\alpha$ \footnote{Greek indices $\alpha,\beta=1,2$ are used to denote vector components in the local particle frame, while Roman indices $i,j=x,y$ refer to vector components in the fixed Cartesian frame.}.  Deformations of interactions between bound particles (i.e.\ 2 per bound edge) are modeled by Hookean elastic springs between specific binding sites, with stiffnesses parameterized by $k_\alpha$ for bonds in the $\ev_\alpha$ direction.  For a given $M \times N$ array of WJ particles with a specific frustration and cohesive stiffness, the elastic ground state is simply computed by energy minimization of the coupled network of springs and rigid particles (see Appendix \ref{app:sim details}).
        
        A second approach to model the elastic ground states is to study the continuum elastic theory of arbitrary rectangular domains of assembled WJ particles.  Such a continuum theory should hold in the limit of large assemblies (i.e.\ $N$ and $M \gg 1$) along with an assumption that the deformations of the ground state from uniform 2D square lattice order, which we take to be the reference state of our theory, are sufficiently small.  In other words, we describe the positions and orientations of WJ particles relative to their corresponding configuration in an {\it unfrustrated} lattice assembly, where they have positions $\rv=x\hat{\xv}+y\hat{\yv}$ and particle frames $\ev_1=\hat{\xv}, \ev_2=\hat{\yv}$.  Upon introducing frustration, particles displace and re-orient as they relax the elastic costs of frustration.  As illustrated in Fig.\ \ref{fig:model}e, we describe the displacement of WJ centers by the field $\uv(\rv)$ and their rotations relative to the fixed 2D Cartesian coordinates by the angle $\theta(\rv)$, i.e.\ the particle frames can be written as $\ev_1=\cos\theta(\rv)\hat{\xv}-\sin\theta(\rv)\hat{\yv}$.  As we describe in the following section, our continuum elastic model is based on a small-$\theta$ expansion of the elastic energy of discrete WJ particle arrays (see Appendix \ref{app:continuum}).

    	\subsection{\label{elasticity}Elasticity of intra-domain distortions}
	        \begin{figure}
	            \centering
	            \includegraphics[width=\columnwidth]{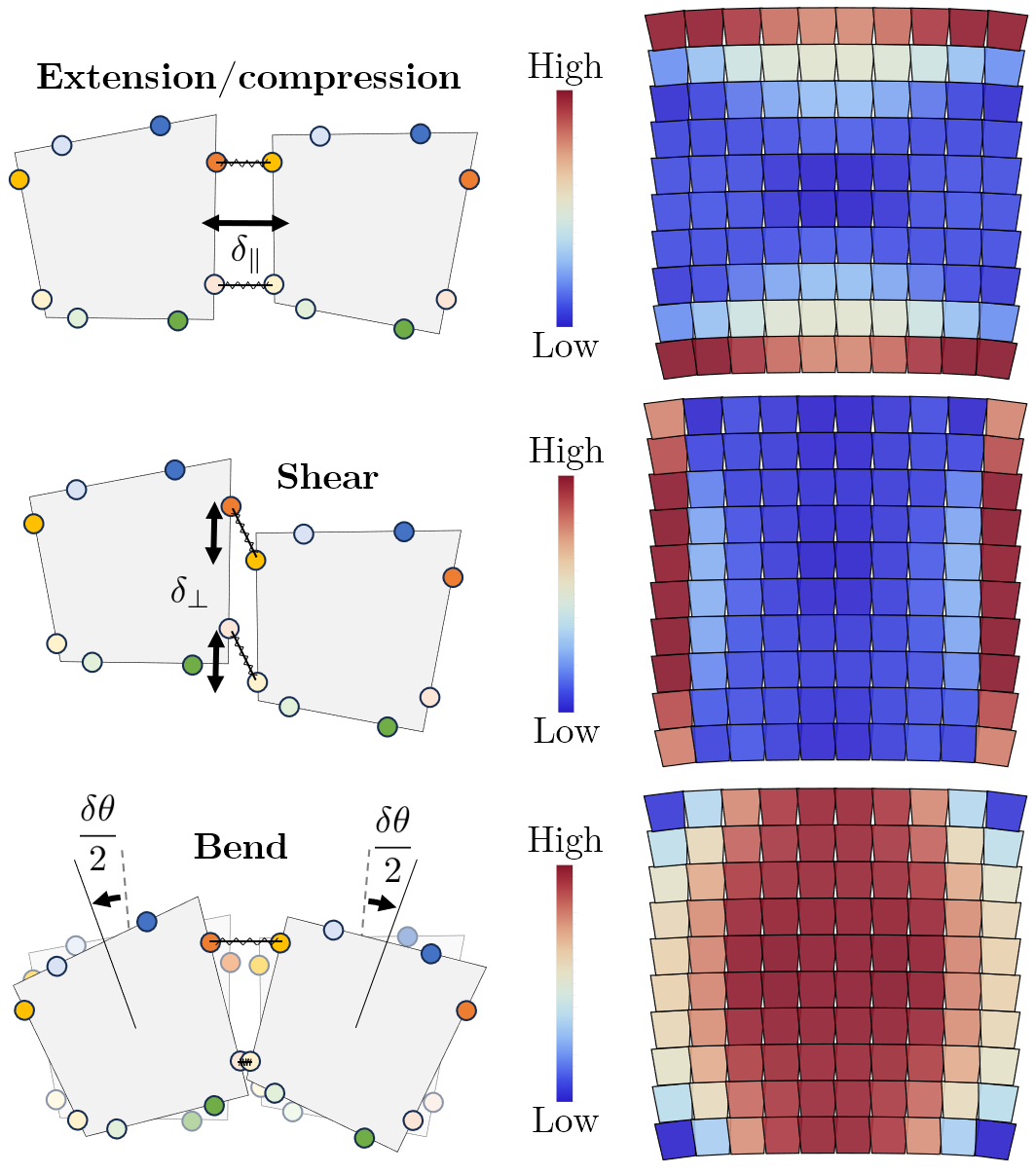}
	            \caption{Three types of elastic deformations between pairs of WJ particles.  Extension or compression occurs when the bound edges of subunits are pulled apart or pushed together by a displacement $\delta_{\parallel}$.  Shear occurs when the bound edges of subunits slide against each other by a displacement $\delta_{\perp}$.  Bending occurs when bound subunits rotate away from their preferred orientation (parallel edges) by a total angle $\delta\theta$.  Corresponding example of a $10\times10$ structure colored by the amount of extension/compression, shear, and bend are shown on the right for isotropic elastic constants $k_1=k_2$.}
	            \label{fig:elastic modes combined}
	        \end{figure}
	        Next, we briefly introduce the mechanical modes of intra-domain deformation, relate the corresponding moduli to microscopic features of the frustrated WJ particles, and finally formulate the continuum elastic energy of finite WJ domains.  
	    
	        As shown schematically in Fig.\ \ref{fig:elastic modes combined}, deformations of inter-particle order can be parameterized by three types of elastic modes: inter-particle stretching, shearing, and bending.  The elastic cost of these modes is controlled by the geometry and stiffness of elastic bonds between bound particles.  At the pairwise level, we can describe the elastic energy of a pair of bound edges along the $\ev_{\alpha}$ direction as
	        \begin{equation}
	            E_{\alpha}=\frac{Y_{\alpha \parallel}}{2}\delta_\parallel^2+\frac{Y_{\alpha\perp}}{2}\delta_{\perp}^2+\frac{B_{\alpha}}{2}(\delta \theta)^2 , 
	        \end{equation}
	        where $\delta_{\parallel}$, $\delta_{\perp}$ and $\delta\theta$ parameterize the respective stretch, shear, and bend displacements (labeled in Fig.\ \ref{fig:elastic modes combined}, left).  The elasticity of these respective modes are determined by stiffnesses and placement of the inter-WJ bonds (see Appendix \ref{app:continuum}) and are given by
	        \begin{equation}
	            Y_{\alpha\parallel}=Y_{\alpha\perp} =2k_{\alpha}; \ B_{\alpha}=\frac{k_{\alpha}v^2}{2},
	        \end{equation}
	        where $v \leq a$ is the distance between linkers along a particle edge (see Fig.\ \ref{fig:model}ac) and $k_{\alpha}$ is the elastic bond stiffness along the $\ev_{\alpha}$ direction.  As an example of the complex interplay between these deformation modes in frustrated WJ domains, we show in Fig.\ \ref{fig:elastic modes combined} the simulated elastic ground state of a $M \times M$ square domain, illustrating the spatial distributions of the three elastic deformation modes for a case $\psi=0$ or polarity along the ``vertical" $\ev_2$ direction.  We return to a detailed analysis of these patterns below, but upon initial inspection it is clear that shearing is largest along left-right free edges, compression/stretching is largest along the bottom-top free edges, and (un)bending of rows away from their preferred curvature is largest in the core of the domains.  
	
	        To understand these emergent patterns of intra-domain distortion, their dependence on domain shape, and their impact on assembly thermodynamics, we study the ground states of a continuum elastic energy of WJ particle assemblies.  This continuum theory was also derived in \cite{Spivack:2022}, but only studied for a limited set of ``tall" finite-width ribbon domains with isotropic binding constants.  As shown in the Appendix \ref{app:continuum}, the elasticity of WJ domains in the continuum limit is described a combination of positional and orientational strains.  Positional strains, resulting from inter-particle stretching and shearing, are measured by the 2D tensor
	        \begin{equation}
	        	\gamma_{ij} (\rv) = \partial_i u_j - \big[R_{ij} (\theta) - \delta_{ij}\big],
	        	\label{eq:strain}
	        \end{equation}
	        where the matrix $R_{ij}(\theta)$ describes 2D rotations by an angle $\theta$ and $\delta_{ij}$ is the Kronecker delta. The form of the strain is invariant under simultaneous rigid rotations of particle centers and orientations, ensuring global rotational invariance of the elastic energy.  Orientational strains are measured by gradients of particle rotation relative to preferred row bending
	        \begin{equation}
	        	\bev(\rv)=\nabla\theta-\kappa_0\Tv,
	        \end{equation}
	        where $\Tv$ describes the preferred direction of 2D inter-particle rotations in the local particle frame.  In the analysis below, we linearize the theory for small frame rotations away from the uniform reference state, resulting in
	        \begin{equation}
	            \gamma_{ij}(\rv)\simeq\partial_i u_j-\theta(\rv)\epsilon_{ij}; \  \bev(\rv)\simeq\nabla\theta-\kappa_0\Tv_0,
	        \end{equation}
	        where $\epsilon_{ij}$ is the 2D anti-symmetric Levi-Cevita tensor and $\Tv_0=-\cos\psi\hat{\xv}+\sin\psi\hat{\yv}$ is the (uniform) approximation of the preferred rotation rate in the small $\theta(\rv)$ limit.
	
	        The elastic energy is computed as an area integral over a rectangular domain ${\cal D}$ (the undistorted reference state)
	        \begin{multline}
	            E_{{\rm elas}}=\frac{1}{2} \int_{{\cal D}}dA\left[Y_x \gamma_{xx}^2(\rv)+Y_y \gamma_{yy}^2(\rv)\right.\\
	            \left.+Y_x\gamma\varepsilon_{xy}^2(\rv) +Y_y\gamma_{yx}^2(\rv)+B_x \beta_x^2(\rv)+B_y\beta_y^2(\rv) \right],
	            \label{eq:elastic energy}
	        \end{multline}
	        where in the small-$\theta$ approximation we neglect the difference between $\{\ev_1,\ev_2\}$ and $\{{\hat{\xv},\hat{\yv}}\}$, thereby identifying the corresponding coordinate indices as $(\alpha = 1) \to (i = x)$ and $(\alpha = 2) \to (i = y)$ and use the fact that the shear and stretch moduli are equal for the WJ model, that is,  $Y_{\alpha\parallel}=Y_{\alpha\perp}  $.  For a given domain, the elastic energy ground state satisfies a set of Euler-Lagrange equations corresponding to force balance
	        \begin{equation}
	        	Y_x\partial_x^2u_x+Y_y\partial_y^2 u_x=-\partial_y\theta; \  Y_x\partial_x^2u_y+Y_y\partial_y^2u_y=\partial_x\theta,
	        	\label{eq:force balance}
	        \end{equation}
	        and torque balance
	        \begin{equation}
	            B_x\partial_x^2\theta+B_y\partial_y^2\theta=-Y_x(\partial_xu_y-\theta)+Y_y(\partial_yu_x+\theta).
	            \label{eq:torque balance}
	        \end{equation}
	        Note that the coupling between displacements and rotations in mechanical equilibrium reveals the mechanism of frustrated elasticity.  Preferred nonzero row curvature, which is favored by the taper angles of the WJ particles, results in nonzero rotation gradients $\nabla\theta\ne0$ and consequently generates a source field for positional strain gradients in Eq.\ (\ref{eq:force balance}).  At the same time, rotations of the local WJ lattice directions away from the particle frames generates a source for orientational strain gradients in Eq.\ (\ref{eq:torque balance}).  Boundary conditions at free edge of finite domain $\partial {\cal D}$ require vanishing forces
	        \begin{multline}
	            \big[Y_x(\partial_xu_x)n_x+Y_y (\partial_yu_x+\theta)n_y\big]_{\partial{\cal D}}=\\
	            \big[Y_x(\partial_xu_y+\theta)n_x+Y_y(\partial_yu_y)n_y \big]_{\partial{\cal D}}=0,
	            \label{eq:force BC}
	        \end{multline}
	        and torques 
	        \begin{equation}
	            \left[B_x(\partial_x\theta-\kappa_0 T_{0,x})n_x+B_y(\partial_y\theta-\kappa_0T_{0,y})n_y\right]_{\partial{\cal D}}=0,
	            \label{eq:torque BC}
	        \end{equation}
	        where $\nv$ describes the normal to domain boundary.
	        
	        The equilibrium equations (Eqs.\ (\ref{eq:force balance}) and (\ref{eq:torque balance})) and boundary conditions (Eqs.\ (\ref{eq:force BC}) and (\ref{eq:torque BC})) constitute to set of coupled, linear PDEs that can be solved by a series expansion (see Appendix \ref{app:solution}).  Non-trivial and spatially-inhomogeneous solutions are required by the torque free boundary conditions in Eq.\ (\ref{eq:torque BC}) which produce nonzero $\theta$ gradients (i.e.\ row bending) at free edges where $\nv\cdot\Tv\neq0$ and vanishing $\theta$ gradients where $\nv\cdot\Tv=0$ (i.e.\ no column bending).

		\subsection{\label{thermo}Thermodynamics of finite-size domains}
	    	We consider a model of the assembly thermodynamics of WJ domains that is determined by the interplay between the elastic costs of frustration and the cohesive gains of binding new particles to the free boundaries of finite domains.  Such an approximation obviously neglects effects of other possible low-symmetry configurations (e.g.\ defective) as well as entropic effects that are present at finite temperature.  Nevertheless, the purpose is to capture the predominant morphologies that emerge in regimes where nearly all subunits belong to assembled domains, i.e.\ super-saturated conditions.  In this case, assembly favors morphologies that minimize the {\it per-subunit free energy of assembly}, which includes the (free-)energetic gains and costs of forming multiunit structures \cite{Hagan:2021}.  In our discrete WJ assemblies, the assembly free energy per subunit for a $M\times N$ domain takes the form
	        \begin{equation}
	            f(M,N)=-(\Sigma_x+\Sigma_y)+\frac{\Sigma_x}{M}+\frac{\Sigma_y}{N}+\mathcal{E}_{\rm ex}(M,N).
	            \label{eq:free energy landscape}
	        \end{equation}
	    	The first term is the constant cohesive gain for bound particles within the bulk, parameterized by attractive binding per edge $-\Sigma_x$ and $-\Sigma_y$ along respective horizontal rows and vertical columns.  The second and third terms together account for the per-subunit cost of unbound free edges on the vertical and horizontal sides of the domain.  The final term $\mathcal{E}_{\rm ex}(M,N)$ denotes the {\it excess energy} and represents the elastic costs of frustration per subunit, computed via the minimization of the spring network energy in our discrete model.  We can similarly use our continuum theory to compute the elastic energy, in this case for a domain of horizontal and vertical dimensions, $W=Ma$ and $H=Na$ respectively.  In this case, the excess energy cost is determined by the mechanical equilibrium of Eq.\ (\ref{eq:elastic energy}) for $W \times H$ domain,
	    	\begin{equation}
	    		\mathcal{E}_{\rm ex}(W,H) = \frac{a^2}{W H} E^*_{\rm elas}(W,H),
	    	\end{equation}
	   	 	where the $*$ denotes the solution of the Euler-Lagrange equations.
	
	    	Inspection of Eq.\ (\ref{eq:free energy landscape}) reveals the basic physical mechanism of the size-dependent thermodynamics of frustrated WJ domains.  On one hand, the cost of open boundaries thermodynamically favors growing all dimensions of an assembly to arbitrarily large sizes, in order to minimize the fraction of unbound edges at free edges of the domains.  On the other hand, the elastic cost of a frustrated WJ domain grows superextensively with domain size \cite{Grason:2016}, meaning that it grows faster than the increase in subunit number in the domain.  This implies that in certain regimes, the competition between boundary and frustration costs selects at least one of the two assembly dimensions to be finite.  In the following sections we first analyze the thermodynamic assembly landscape resulting from this interplay for the particular case of $\psi=0$ polarization before then considering a broader range of misfit polarities.

    \section{\label{sec:psizero} Alignment of misfit polarization and lattice anisotropy:  $\psi = 0$ case }

		Here, we analyze the case of $\psi=0$ where misfit polarity is aligned to a 2D assembly direction such that $\boldsymbol{\kappa}_0$ is parallel to $\ev_2$ and the ``horizontal'' rows of the WJ domain favor curvature.  We first study the mechanics and elastic energy of inhomogeneous ground states as function of domain shape and then describe the thermodynamics optimal morphologies.

		\subsection{Elastic energy landscape: $\psi = 0$}
\begin{figure*}
                \centering
                \includegraphics[width=\textwidth]{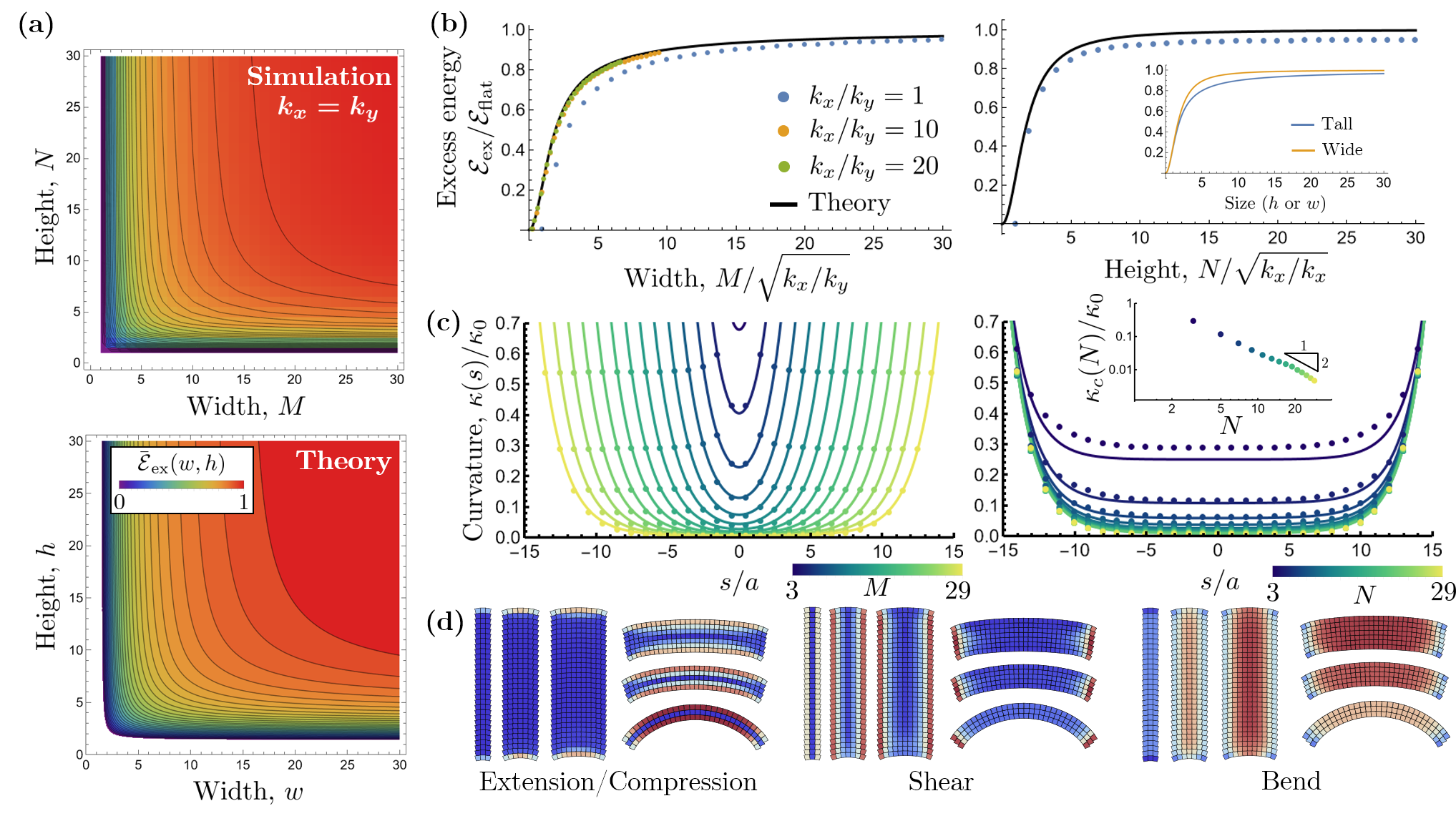}
                \caption{Energetics and shape of WJ particle assemblies.  \textbf{a)} Excess energy landscapes for $M\times N$ assemblies from simulations and continuum theory.  Simulations have isotropic elastic constants with $k_x=k_y$.  \textbf{b)} Excess energy of tall assemblies with fixed height $N=30$ and varying width $M$ (left) and wide assemblies with fixed width $M=30$ and varying height $N$ (right).  Inset compares the excess energy of tall and wide assemblies.  \textbf{c)} Centerline curvatures as a function of position $s$ along an arc for tall $Mx30$ (left) and wide $30xN$ (right) assemblies.  Inset shows the height dependence of the curvature.  \textbf{d)} Example assemblies highlighting the degree of extension/compression, shear, and bend in tall and wide assemblies.}
                \label{fig:excess energy}
            \end{figure*}

       		The detailed solution of the continuum theory (Eqs.\ (\ref{eq:force balance}) and (\ref{eq:torque balance})) is presented in Appendix \ref{app:solution}.  For this particular polarization, the free boundary conditions (Eqs.\ (\ref{eq:force BC}) and (\ref{eq:torque BC})) are such that the curvature of WJ rows achieves its preferred magnitude $\kappa_0$ on the vertically oriented sides of the domains, while its value in the interior of the domain may deviate from $\kappa_0$ depending on the domain size.  The relaxation derives from a competition between preferred row bending and accumulated costs of either intra-row stretching or inter-row shears, and is thus characterized by ``bend penetration" length scales determined by the ratios between elastic constants,
            \begin{align}
                \lambda_x&=\sqrt{B_x/Y_y};\ \lambda_y=\sqrt{B_x/Y_x}.
            \end{align}
            In Appendix \ref{app:approximation}, we introduce an approximation based on observations made on the structures (e.g.\ Fig.\ \ref{fig:excess energy}d) that allows us to determine the elastic ground state energy of $W\times H$ domains, which is given in closed form by
            \begin{multline}
                \tilde{\mathcal{E}}_{\rm ex}(\psi=0;w,h)=\frac{\mathcal{E}_{\rm ex}}{\mathcal{E}_{\rm flat}}\\
                =1-\frac{2}{w^2}\sum_{m\textrm{ odd}}\frac{1}{\frac{m^2\pi^2}{4w^2}+1-\frac{2w}{m\pi h}\tanh\frac{m\pi h}{2w}},
                \label{eq:excess energy density}
            \end{multline}
            where $\mathcal{E}_{\rm flat}=\frac{1}{2}B_x\kappa_0^2 a^2= $ is the (per particle) {\it flattening cost} to completely unbend the rows from their preferred curvature, and the rescaled width and height are $w=W/(2\lambda_x)$ and $h=H/(2 \lambda_y)$, respectively.  Notably, it can be easily verified that the second term in Eq.\ (\ref{eq:excess energy density}) vanishes in the double limit of $w \to \infty$ and $h\to \infty$, which shows that the bulk state is a uniformly flattened, except at the free boundaries.  In general the excess energy approaches this upper limit $\mathcal{E}_{\rm flat}$ with increasing size.  It is interesting to note that in the limit of asymptotically small assemblies $w\ll1$ and $h\ll1$, the excess energy takes on the form (see Appendix \ref{app:approximation})
            \begin{equation}
                \tilde{\mathcal{E}}_{\rm ex}(\psi=0;w\ll1,h\ll1)\simeq\frac{1}{3}\sum_{m,n\textrm{ odd}}\frac{1}{m^2n^2\left(\frac{m^2}{w^2}+\frac{n^2}{h^2}\right)},
            \end{equation}
            which is symmetric about $h=w$, indicating that elastic anisotropy only enters beyond quadratic order in domain sizes.

            Fig.\ \ref{fig:excess energy} shows a comparison of excess energy from simulations of $M\times N$ assemblies of discrete subunits with continuum theory of $W\times H$ (or $Ma\times Na$) domains, showing qualitative and quantitative agreement notwithstanding the small-$\theta$ approximation made in the continuum formulation.  The excess energy landscapes show two zero energy states corresponding to single rows (wide) and columns (tall).  As the width of the tall assemblies and the height of the wide assemblies increases, there is an accumulation of strain due to the frustration or shape misfits, leading to a super-extensive growth in excess energy.  The excess energy for tall (fixed height and varying width) and wide (fixed width and varying height) assemblies are shown in Fig.\ \ref{fig:excess energy}b, left and right, respectively.  In general, these show that the excess energies increase monotonically with both the width and height of the domains, albeit with obvious differences in their rate of increase.  In particular, the excess energy of tall assemblies accumulates more gradually than that of the wide assemblies.

            In Fig.\ \ref{fig:excess energy}cd, we show the mechanical equilibria of the two classes of anisotropic morphologies,  tall ($H \gg W$) and wide ($W\gg H$) ribbon-like domains, from discrete particle simulations comparing their distinct intra-domain mechanics.  First, we note that away from the boundaries of either type of domain, elastic ground states relax to a pattern that is effectively uniform along longer dimension.  Next, we observe that tall and wide morphologies both continuously approach the flattened state (i.e.\ uniformly unbent rows) as their narrow dimension grows, but they do so via distinct patterns.  Tall ribbons exhibit large shears between rows at their free edges over a boundary zone of size $\sim\lambda_x$, with rows nearly fully unbending away from this boundary zone.  In contrast, flattening in wide ribbons is mediated by a competition between bending and intra-row compression, with shears vanishing along most of their long dimension.  Unlike shears in tall ribbons, the differential compression along rows in wide ribbons is not confined to a finite boundary layer, and instead extends through the thickness.  Another key distinction is the fact that row bending at the center of the wide ribbons remains non-zero and continues to decrease as the ribbons grow in their narrow dimension, whereas row bending in tall ribbons falls to zero away from the shear boundary layer.  Fig.\ \ref{fig:excess energy}c shows this contrasting behavior for row curvature behavior along the central horizontal row of tall and wide ribbons, with the former relaxing to uniformly unbent rows away from the boundary zone while the latter adopting a non-zero value that continuously unbends with increasing thickness.

            These distinct mechanics are captured in the analytical forms of the elastic energy in the infinite width or height limits of the continuum solutions.  The two morphologies of interest are those that grow tall with $H\gg W$ or wide with $W\gg H$.  In the limit of $H/W\rightarrow\infty$, the excess energy of tall ribbons simplifies to
            \begin{equation}
                \tilde{\mathcal{E}}_{\rm ex}^{(\rm tall)}(w)\equiv\tilde{\mathcal{E}}_{\rm ex}(w,h\rightarrow\infty)=1-\frac{\tanh w}{w},
                \label{eq: tall}
            \end{equation}
            which was derived previously in Ref.\ \cite{Spivack:2022}.  Notably, the exponential form of the width dependence is a consequence of the finite shear boundary at the sides of the ribbon.   That is, the boundary conditions required that the curvature at the free edges of the domain achieve their preferred value, but as one moves away from the free edges, the resistance to inter-row shear exponentially screens the curvature beyond a distance comparable to $\lambda_x$.  Hence, local arrangements within the core of the domain become effectively insensitive to size for tall ribbons with $W \gg \lambda_x$. This is consistent with the asymptotic form $\tilde{\mathcal{E}}_{\rm ex}^{(\rm tall)}(w\gg1) \simeq 1-1/w$, indicating that the entire width of the ribbon is unbent, with the exception of a finite boundary layer proportional to $\lambda_x$ where the elastic penalty of inter-row shearing is consequently lower.

            In the limit of wide ribbons $W/H\rightarrow\infty$, the excess energy takes on the form
            \begin{equation}
                \tilde{\mathcal{E}}_{\rm ex}^{(\rm wide)}(h)=\tilde{\mathcal{E}}_{\rm ex}(w\rightarrow\infty,h)=\frac{h^2}{3+h^2},
                \label{eq: wide}
            \end{equation}
            which corresponds to nearly uniformly-curved (i.e.\ circular) ribbons characterized by a mid-row curvature
            \begin{equation}
                \kappa^{(\rm wide)}_c(h)=\frac{\kappa_0}{1+h^2/3}.
                 \label{eq: kappawide}
            \end{equation}
            A detailed calculation for this can be found in Appendix \ref{app:mesoscopic curvature}.  A key distinguishing feature of the wide ribbons is their {\it mesoscale curvature} which varies with the finite thickness of the domain itself.  This mesoscale curvature results in a large size scale since $\kappa_c(h)^{-1} \geq \kappa_0^{-1}$ and we generally consider the limit of small curvature frustration where $\kappa_0H\ll1$.  To clarify, due to the finite curvature ``wide'' ribbons, it is not strictly possible to extend $W$ beyond $2\pi \kappa^{(\rm wide)}_c(h)^{-1}$, as the free ends ribbons will meet.  However, we expect that in the small frustration limit (i.e.\ $\kappa_0a\ll1$) the elastic energy of the small-$\theta$ expansion accurately describes the circumferentially-uniform strain distributions in closed annular ribbons.  Notably, the thickness dependence of row-curvature at the core of domains is power-law as opposed exponential, with $\kappa_c(h)\sim h^{-2}$ indicating that even for $h \gg 1$, residual curvature propagates to the center of the domain and is not screened by a free boundary (Fig.\ \ref{fig:excess energy}c, inset).  While for narrow assemblies (i.e.\ $w\ll1$ or $h\ll1$), both types of assemblies exhibit the same quadratic growth of excess energy with narrow dimension ($\tilde{\mathcal{E}}_{\rm ex}^{(\rm tall)}(w\ll1)\simeq w^2/3$ and $\tilde{\mathcal{E}}_{\rm ex}^{(\rm wide)}(h\ll1)\simeq h^2/3$), the lack of boundary screening in wide-ribbon assemblies leads to distinct and more rapid power-law approach the asymptotically flattened limit $\tilde{\mathcal{E}}_{\rm ex}^{(\rm wide)}(h \gg 1) \simeq 1-3/h^2$.  
        
            The analytic forms of excess energy for tall and wide limits are shown as black lines in Fig.\ \ref{fig:excess energy}b and are compared in the inset, highlighting the respective similarity and distinctions in the narrow and thick ribbon limits.  We next consider the thermodynamic consequences of those differences in the thickness dependence of excess energy for self-limitation.
        
        \subsection{Self-limiting domain size selection: $\psi = 0$ case}
            \begin{figure*}
                \centering
                \includegraphics[width=\textwidth]{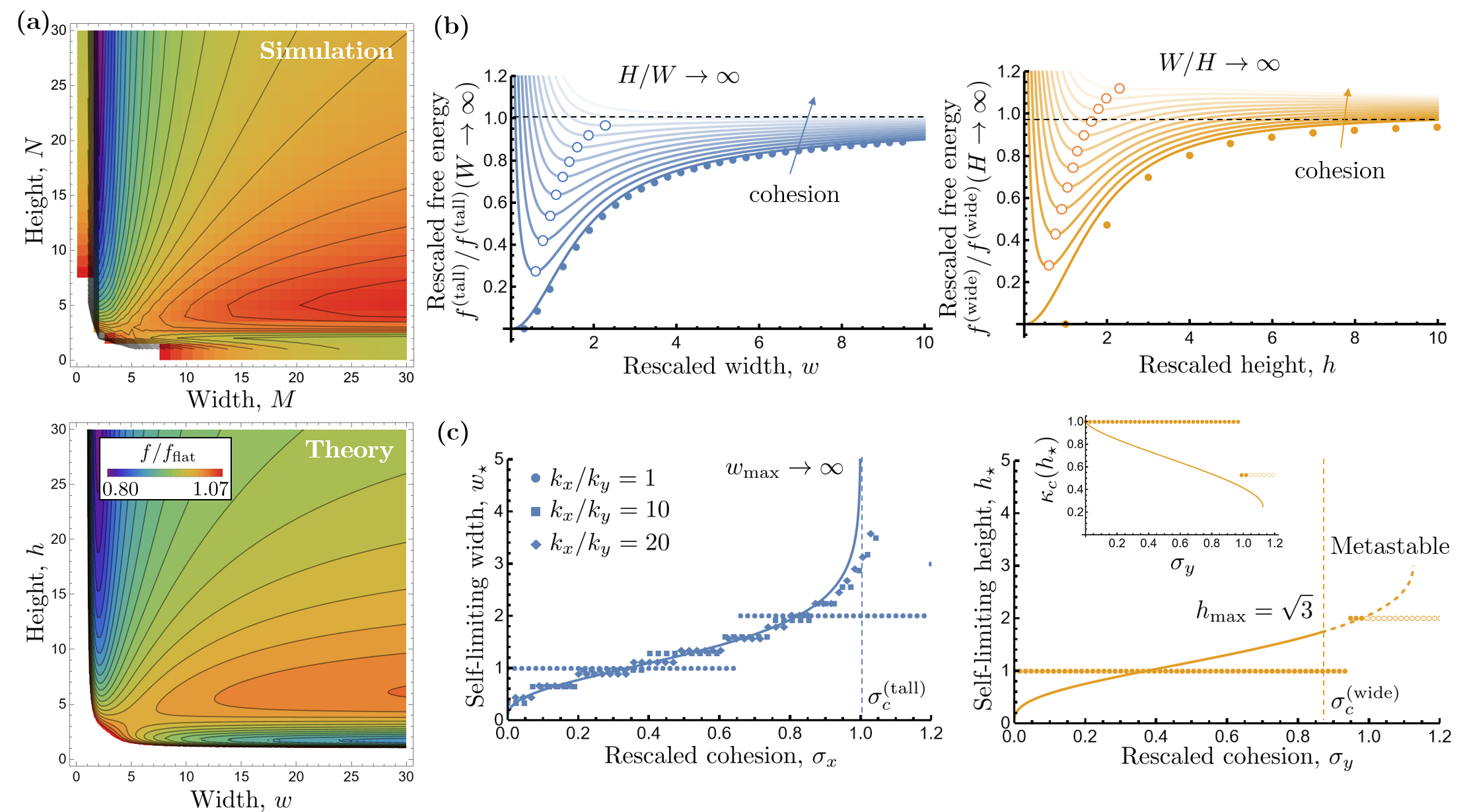}
                \caption{Assembly free energy and self-limiting sizes of WJ particles assemblies.  \textbf{a)} Free energy landscapes for $\Sigma_x/\mathcal{E}_{\rm flat}=\Sigma_y/\mathcal{E}_{\rm flat}=0.8$ with isotropic elasticity $k_x=k_y$ for simulations and $\sigma_x=\sigma_y=0.8$ for theory.  \textbf{b)} Free energies for tall (left) and wide (right) assemblies for varying cohesion.  Solid dots are simulations of $M\times100$ and $100\times N$ assemblies with $k_x/k_y=10$.  Empty circles indicate positions of the local minima with dashed lines showing the flattening energy.  \textbf{c)} Self-limiting sizes of tall (left) and wide (right) assemblies as functions of cohesion.  Inset shows the centerline curvature of wide assemblies.}
                \label{fig:self-limiting size}
            \end{figure*}

            We now consider the consequences of anisotropy in the excess energy landscapes on the selection of self-limiting domain dimensions arising from the competition between the cohesive costs at the boundaries and the intra-domain elastic cost due to frustration.  These self-limiting domains are described by minima in the per subunit assembly free energy, Eq.\ (\ref{eq:free energy landscape}).  We first consider the case of isotropic binding strengths along orthogonal directions, $\Sigma_x=\Sigma_y$, and plot the per-subunit assembly free energies for both the discrete and continuum models in Fig.\ \ref{fig:self-limiting size}a.  As can be seen, there are two low-energy channels through which the assembly free energy can decrease: one for fixed $W$ and $H\rightarrow\infty$ corresponding to tall ribbons and another for fixed $H$ and $W\rightarrow\infty$ corresponding to wide (annular) ribbons.  Broader considerations of the $\psi=0$ assembly free energy landscapes for arbitrary values of $\Sigma_x$ and $\Sigma_y$ reveals that these two ribbons states, i.e.\ tall (extending along $\ev_2$) or wide/annular (extending along $\ev_1$), are the only possible minima.  In addition, these low energy states assemble unlimited along their long dimensions while the narrow dimensions may be finite depending on the value of the cohesive energy of their free edges.  The thermodynamics of the narrow dimensions are determined by the (per subunit) assembly free energy,
            \begin{subequations}
                \begin{align}
                    \frac{f^{(\rm tall)}(w)}{f^{(\rm tall)}(\infty)}&=\frac{\sigma_x}{w}+1-\frac{\tanh w}{w},\\
                    \frac{f^{(\rm wide)}(h)}{f^{(\rm wide)}(\infty)}&=\frac{\sigma_y}{h}+\frac{h^2}{3+h^2},
                \end{align}
            \end{subequations}
            where $\sigma_x$ and $\sigma_y$ are rescaled edge binding energies given by
            \begin{subequations}
                \begin{align}
                    \sigma_x&=\frac{\Sigma_x}{B_x\kappa_0^2a\sqrt{B_x/Y_y}}=\frac{\Sigma_x}{B_x\kappa_0^2a\lambda_x},\\
                    \sigma_y&=\frac{\Sigma_y}{B_x\kappa_0^2a\sqrt{B_x/Y_x}}=\frac{\Sigma_x}{B_x\kappa_0^2a\lambda_y}.
                \end{align}
            \end{subequations}
            These rescaled binding energies can be interpreted as the ratio of the cohesive cost of having free boundaries to cost of flattening the boundary of a ribbon over the bend-penetration lengths $\lambda_x$ or $\lambda_y$ for tall or wide ribbons, respectively.  The free energies of tall (blue) and wide (orange) assemblies are plotted in Fig.\ \ref{fig:self-limiting size}b along with the local minima indicated by the empty circles.  The self-limiting sizes can be determined by locating the local minimum $w_{\star}$ of $f^{(\rm tall)}(w)$ and $h_{\star}$ of $f^{(\rm wide)}(h)$ for given binding strengths $\sigma_x$ and $\sigma_y$, plotted in Fig.\ \ref{fig:self-limiting size}d.  In general, the self-limiting sizes increase with edge binding strength, which can be understood by a simple consideration of the small cohesion and consequently small self-limiting size limit of the continuum theory.  In this limit, the balance between the quadratic growth of excess energy with size and the inverse dependence of edge cost with size leads to the power laws $w_* \simeq ( 3 \sigma_x/2)^{1/3}$ and $h_* \simeq ( 3 \sigma_y/2)^{1/3}$, which also shows that self-limiting domain sizes decrease with frustration as $\sim \kappa_0^{-2/3}$ in this weak cohesion regime.
    
            Beyond this weak cohesion and narrow thickness limit, tall and wide ribbons exhibit distinct thermodynamic dependence of finite domain thickness on cohesion, most notably at the self-limiting to bulk transitions.  As described previously \cite{Hagan:2021}, the nature of the self-limiting to bulk transition depends on the specific form of the saturation of excess energy in the limit of large domain sizes.  As described in the previous section, the elastic energy of tall and wide ribbons approach that of bulk state with leading corrections  $-1/w$ and $-1/h^2$ respectively, differences that characterize a relatively more gradual vs.\ rapid expulsion of gradients in the ground states for increasingly larger domain thicknesses.
            
            Tall ribbons exhibit a second-order transition between finite-width ribbons and bulk states in which $w_*$ increases continuously with $\sigma_x$ up to critical value $\sigma_c^{({\rm tall})}=1$ where it diverges continuously.  This is due to the fact that as the edge cohesive strength is increased, the local minimum of the free energy for tall assemblies remains below the energy of the bulk flattened ($w\rightarrow\infty$) state and approaches that state continuously as $\sigma_x \to \sigma_c^{({\rm tall})}$ (Fig.\ \ref{fig:self-limiting size}b, left).  In Fig.\ \ref{fig:self-limiting size}c, we compare the analytical prediction of the self-limiting sizes from continuum theory to those obtained from the discrete WJ simulations for a range of values of elastic anisotropy $k_x/k_y$.  Notably, this ratio controls the size of the bend-penetration $\lambda_x$ relative to the discrete WJ particle size $a$.  As $k_x/k_y$ is increased, row bending becomes increasingly stiffer than inter-row shearing, leading to a much longer range of frustrated curvature propagation.  The effect is that the size scale over which elastic energy accumulates and consequently the self-limiting sizes proportionately increase with the elastic anisotropy $k_x/k_y$ or $\lambda_x$, as seen by the collapse of the data in Figs.\ \ref{fig:excess energy}b and \ref{fig:self-limiting size}c.  This increase in size range results in the tall ribbons exhibiting more distinct, integer values of finite width prior to the divergent region for $\sigma_x \lesssim \sigma_c^{({\rm tall})}$.  While in principle the equilibrium self-limiting width of tall ribbons is predicted to grow arbitrarily large, $w_{\star}\to\infty$, as the strength of cohesion approaches $\sigma_c^{(\rm tall)}$, it is likely that the combined effects of width-fluctuations \cite{Spivack:2022} and the extreme sensitivity of size near $\sigma_c^{(\rm tall)}$ allow any practical control of width in physical experiments. 
            
            Fig.\ \ref{fig:self-limiting size}c shows the corresponding prediction for the growth of the self-limiting thickness $h_*$ of wide assemblies with $\sigma_y$, which exhibits a first-order transition between self-limiting and bulk states.  This means that above the transition value $\sigma_y =\sigma_c^{({\rm wide})}=\sqrt{3}/2$, the local minimum of $f^{(\rm wide)}(h)$ exceeds the free energy of the bulk state, making it metastable.  This metastable branch of self-limiting annular ribbons eventually loses stability to the bulk uniform state above an edge cohesion $\sigma_s^{({\rm wide})}=9/8$.  This first-order transition implies that there is an upper limit to the equilibrium self-limiting thickness of annular ribbons given by $h_*(\sigma_c^{({\rm wide})})=\sqrt{3}$.  Notably, within our discrete WJ particle model studied here, the elastic length scale $\lambda_y=v$ is independent of the elastic interaction stiffnesses and is therefore less than or equal to particle size $a$ since $v \leq a$, which implies that the size of equilibrium self-limiting annular assemblies is comparable to that of the particle dimension $a$ and is therefore comparatively more limited than tall ribbons, whose size selection can vary with $k_x/k_y$.  While not the goal of this study,  it is possible to access a broader range of bend to stretch ratios through additional different microscopic features, for example, by introducing specific {\it repulsive} binding sites on the faces of WJ particles, as described previously \cite{Spivack:2022}.  Notwithstanding the limited range of equilibrium self-limiting thicknesses of wide ribbons in the present model, we compare the dependence of the self-limiting thicknesses $h_{\star}$ with $\sigma_y$ of discrete WJ simulations for annular ribbons to the continuum theory predictions in Fig.\ \ref{fig:self-limiting size}c, right. 
    
            Last we note that predicted increase in self-limiting thickness $h_{\star}$ with cohesion is accompanied by corresponding {\it decrease} in the mesoscopic curvature of the ribbon itself, due to the increasing differential compression within the rows.  The predicted variation of ribbon curvature $\kappa_c (h_*)$ with $\sigma_y$, as well as that obtained from discrete WJ simulations, is plotted in the inset of Fig.\ \ref{fig:self-limiting size}c.  This shows that the mesoscopic radius of curvature varies from the preferred value $\kappa_0^{-1}$ for $\sigma_y \to 0$ up to twice that value at the transition point at $\sigma_y= \sigma_c^{({\rm wide})}$.  Hence, the mesoscopic dimensions of annular ribbon morphologies exhibit distinct thermodynamic sensitivity to inter-particle cohesion.  In particular, changes in the domain thickness at the particle scale manifest in large changes in the gross morphology of wide ribbons at the size scale $\kappa_0^{-1} \gg a$.

        \subsection{Phase diagram: $\psi = 0$ case}
        	\begin{figure}
        		\centering
        		\includegraphics[width=\columnwidth]{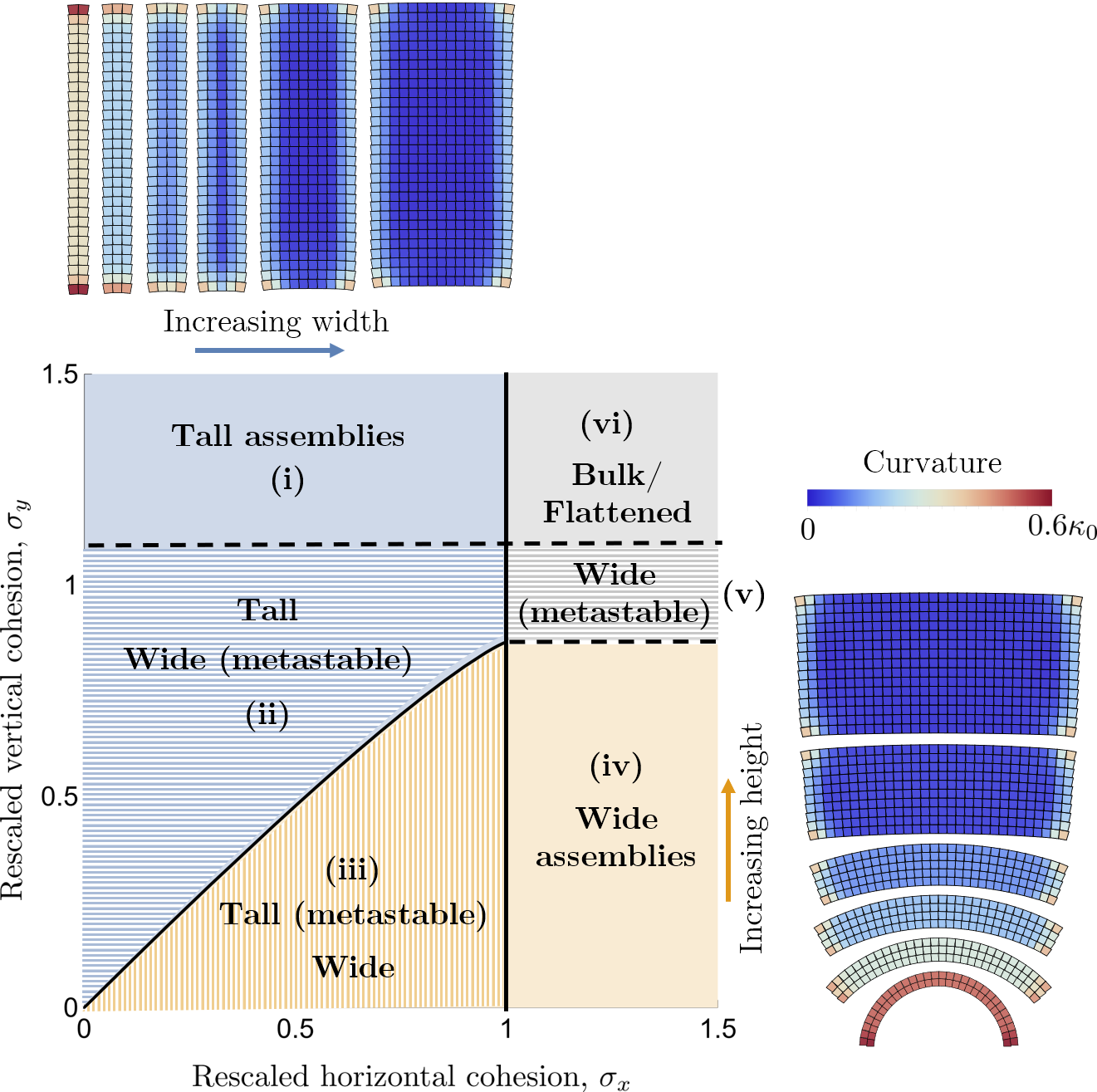}
        		\caption{Phase diagram for tall ($H>W$) and wide ($W>H$) assemblies.  Tall assemblies are energetically stable in blue regions, wide assembles are energetically stable in orange regions, and bulk assemblies are stable in gray regions.  The patterned regions indicate where at least one other structure is metastable.}
        		\label{fig:phase diagram}
        	\end{figure}
            As described above, among the rectangular domains for $\psi =0$ misfit polarization, there exist two classes of minima describing tall and wide ribbons.  These states differ qualitatively in terms of their gross morphology, internal strain distributions, and elastic costs of frustration.  As a consequence, even for the case of equal scaled edge cohesions, $\sigma_x=\sigma_y$, their ground state free energies differ, with the free energy of tall ribbons generically lower than that of wide ribbons, which are therefore metastable for this isotropic binding case.  
    
            As shown in Fig.\ \ref{fig:self-limiting size}b, the ground state free energies $f^{(\rm tall)}(w_{\star})$ and $f^{(\rm wide)}(h_{\star})$ are governed by binding strengths along distinct edge directions $\sigma_x$ and $\sigma_y$, respectively.  Therefore, the thermodynamic selection between these distinct states is influenced by the ratio of binding along distinct directions, reflecting a preference to extend a domain along a low-edge energy face.  The phase diagram of ground state morphologies is shown in Fig.\ \ref{fig:phase diagram}, showing that WJ assembly exhibits polymorphism among three mesoscopically distinct states: tall, wide, and bulk assemblies.
            
            The phase diagram is divided into six main regions, determined by the existence of a local minimum and whether that minimum is the ground state.  Recall that the critical binding strengths at which the local minima of tall and wide assemblies disappear are $\sigma_x=1$ and $\sigma_y=9/8$, respectively.  Therefore, when binding is too strong (i.e.\ $\sigma_x>1$ and $\sigma_y>9/8$), bulk assembly of unlimited flattened 2D structures occurs, as indicated by the solid gray region (vi).  Decreasing the horizontal ($\ev_1$)  binding strength below its critical value ($\sigma_x<1$), results in tall straight ribbons extending unlimited along the vertical ($\ev_2$) binding directions, as indicated by the solid blue region (i).  Lowering the vertical binding strength below its critical value ($\sigma_y<9/8$), a local minimum develops in the free energy of the wide annular ribbons.  For $\sigma_c^{\rm (wide)}<\sigma_y<9/8$, this minimum is metastable since the lowest energy state is still that of tall straight ribbons or bulk assembly, as indicated by the patterned blue (ii) or gray (v) regions.  When the vertical binding strength is sufficiently weak (iii), finite thickness annular ribbons become the equilibrium state while the tall straight ones become metastable.  The equilibrium transition between the two morphologies occurs when $\sigma_y<\sigma_c^{\rm (wide)}$ and $\sigma_y\simeq\sigma_x-\sigma_x^{5/3}/(10\sqrt[3]{12})$, indicating that for equal binding strengths $\sigma_x=\sigma_y$, tall ribbons are generally favored.  That is, in the absence of cohesive anisotropy, the polarization of misfit and its elastic effects bias the thermodynamics towards mesoscopically tall straight ribbons, owing to the relatively ``softer'' accumulation of elastic energy of tall ribbons with width.  Despite this, we note that only a relative small bias in binding energy along the horizontal and vertical directions is needed to stabilize the otherwise metastable wide and mesoscopically curved ribbons.  
    

    \section{\label{sec:psinonzero}Misfit polarity along off-lattice directions:  $0<\psi<\pi/4$ }
        \begin{figure}
            \centering
            \includegraphics[width=\columnwidth]{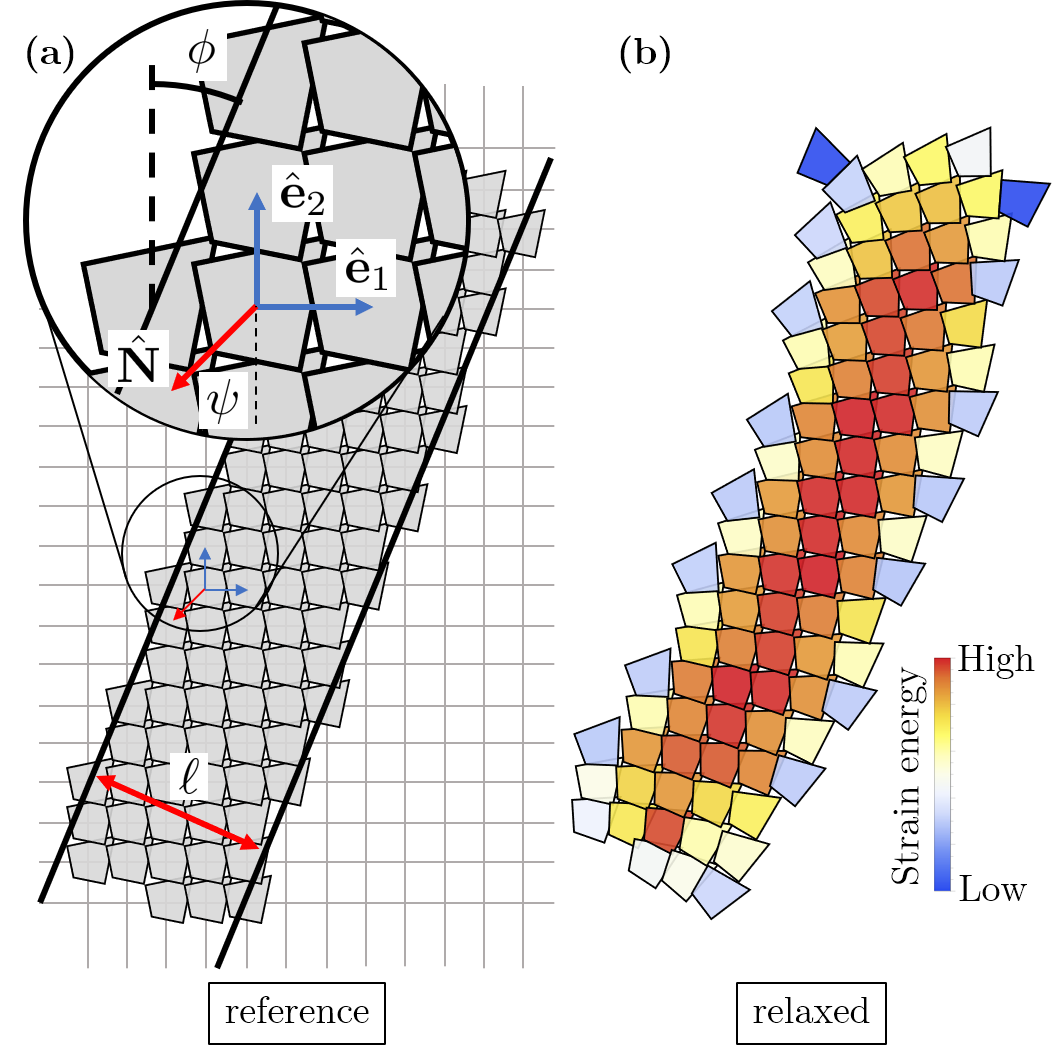}
            \caption{Assembly along off lattice directions.  \textbf{a)} Definition of the angle $\phi$ between the long-axis of a ribbon and the $\ev_2$ binding direction in the reference state.  \textbf{b)} Relaxed assembly for off-lattice direction $\phi=\pi/8$ and shape asymmetry $\psi=\pi/4$, with curvatures $\kappa_xa=\kappa_ya=0.405$.}
            \label{fig:phi_definition}
        \end{figure}

        In Sec.\ \ref{sec:psizero}, we found that the emergent morphologies of frustrated WJ assembly derive from two sources of anisotropy: i) the binding strengths along the two binding directions $\ev_1$ and $\ev_2$ and ii) the polarization of the WJ shape misfit itself.  In this section we consider a more general class of ribbon morphologies where the misfit polarization $\kapv_0$ is not aligned to one of the two lattice directions ($\ev_1$ or $\ev_2$), i.e.\ the cases where $0<\psi<\pi/4$.  Note that $\psi = \pi/4$ represents the largest possible deviation from a nearest-neighbor binding direction since $\psi<0$ or $\psi>\pi/4$ simply correspond to mirror reflections of the WJ particles and hence do not change any results.  For the $\psi = 0$ case, we found that the elastic cost of frustration is lowest for tall ribbons, that is, when assembly grows arbitrary long along the misfit polarization direction.  In that case, the bias of elastic energy towards forming tall ribbons is compatible with the bias of low-edge energies towards forming 2D rectangular domains, which are lowest when the direction of free edges is aligned with a nearest-neighbor binding direction.  More generally, assembly of finite-width ribbons could occur in off-lattice directions, a possibility which we characterize by introducing the angle $\phi$ between the long-axis of the ribbon and the $\ev_2$ binding direction in the assembly, which we show schematically in Fig.\ \ref{fig:phi_definition}.  Notably for $\phi \neq n \pi/2$ (where $n$ denotes an integer), free edges are ``terraced'' leading to a greater cohesive cost per unit length of the ribbon compared to the flat edges of rectangular domains discussed in Sec.\ \ref{sec:psizero}, a microscopic effect underlying facet formation more generally found in crystal assembly \cite{Rottman:1984}.  
        
        In this section we analyze the thermodynamic competition between assembly of finite domains along low-edge energy directions of the 2D WJ ``lattice" (i.e.\ $\phi = n \pi/2$) versus assembly along the direction favored by misfit polarization (i.e.\ $\phi = \psi$).  We first consider assembly free energy landscapes for domain edges are aligned to the lattice directions $\ev_1$ and $\ev_2$, but for misfit polarities which are off those lattice directions (i.e.\ $\phi =0$ and $0<\psi<\pi/4$).  We then consider possible ribbon ground states whose free edges are oriented away from the low-edge energy directions (i.e.\ $\phi \neq n\pi/2$).  For simplicity, we restrict our analysis in this section to the case of isotropic elastic parameters, meaning that $k_x=k_y$ and correspondingly $B_x=B_y$, $Y_x=Y_y$ and $\lambda_x=\lambda_y$.
        
        \subsection{Thermodynamic landscapes for $\phi=0$ and $0<\psi<\pi/4$ domains}
            \begin{figure*}
                \centering
                \includegraphics[width=\textwidth]{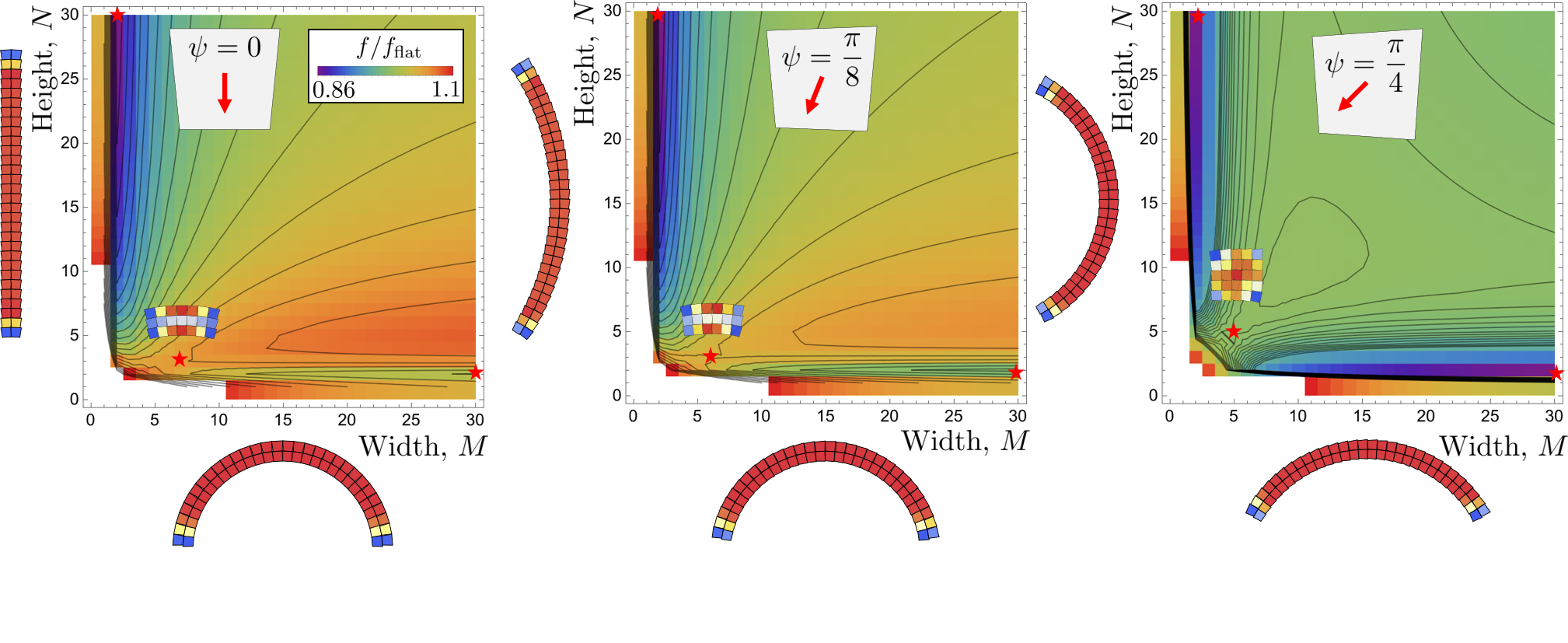}
                \caption{Free energy landscapes for frustration directions $\psi=0,\pi/8,\pi/4$ with fixed frustration $\kappa_0a=0.1$ and isotropic elastic constants $k_x=k_y$.  Example structures shown for the low energy channels in the energy landscapes and the saddle point.}
                \label{fig:polar 0 22 45}
            \end{figure*}
            In Fig.\ \ref{fig:polar 0 22 45} we compare the per-subunit assembly free energy landscapes from discrete WJ simulations with isotropic binding $\Sigma_x=\Sigma_y$ for three cases of misfit polarization, $\psi =0, \pi/8$ and $\pi/4$.  Per the design in Fig.\ 2, these changes in polarization are affected by appropriate changes in the taper angles, $\alpha_x$ and $\alpha_y$, leading to kite-shaped quadrilaterals for $\psi \neq n \pi/2$.  Upon inspection, it is clear that the gross features of the landscape are preserved with changes of polarization.  Namely, the existence of two low-energy basins, oriented vertically and horizontally, representing two classes of finite-thickness ground states whose configurations has one dimension that extends unlimited.  More careful inspection shows that the asymmetry between the vertical and horizontal basins is progressively reduced as the misfit polarization is increased from $\psi=0$.  This change is also reflected in the ground state shapes of the two ribbon classes.  For $\psi =0$, vertical ribbons are mesoscopically straight, while horizontal ribbons are highly curved.   Upon increasing the misfit polarization to $\psi=\pi/8$, vertical ribbons adopt a slight bend, while horizontal ribbon curvature is slightly reduced.  Finally, for maximal off-lattice polarization of $\psi=\pi/4$, the two ribbon morphologies adopt the same mesoscopic curvature.  Indeed, in this case there is clear symmetry in the assembly free energy landscape about the $M = N$ line.  This emergent symmetry in the landscape implies that as $\psi \to \pi/4$ the two ribbon morphologies become degenerate.
    
            This basic dependence of the ground state thermodynamics on polarization direction can be understood simply in terms properties of the solutions of the continuum theory.  From inspection of Eqs.\ (\ref{eq:force balance}) and (\ref{eq:torque balance}), it is clear that positional and orientational strains, $\uv$ and $\theta$ respectively, are linearly coupled to each other as well as to the WJ particle frustration $\kapv_0$ via the boundary condition of Eq.\ \ref{eq:torque BC}, which for isotropic elasticity reduces to $(\nabla \theta - \kappa_0 \Tv) \cdot \nv = 0$ at the free edges of the domain.  Hence, for $\psi =0$ this requires preferred row bending at the vertically oriented free edges and vanishing column bending at the top and bottom of the domains, leading to a solution in which $\uv$ and $\theta$ are linear functions of $\kappa_0$ and a corresponding elastic cost that is quadratic in preferred row curvature $\kappa_0$.  Likewise, the case of $\psi = \pi/2$ corresponds to the same set of solutions just rotated by $\pi/2$, i.e.\ using the results of Eq.\ (\ref{eq:excess energy density}), $\tilde{\cal{E}}_{\rm ex} ( \psi=\pi/2; w, h)=\tilde{\cal{E}}_{\rm ex} ( \psi=0; h, w)$.  For the case of $\psi \neq n \pi/2$, misfit polarity $\Nv$ (and $\Tv$) have projections along both row and column directions resulting in preferred curvatures $\kappa_1= - \kappa_0 \cos \psi$ and $\kappa_2= \kappa_0 \sin \psi$, respectively.  In this case of off-lattice polarization, the solution for the positional strains is simply a linear superposition of the two solutions for $\psi = 0$ (vertical polarization) and $\psi= \pi/2$ (horizontal polarization).  A more subtle, but valuable, feature of this superposition is that the {\it elastic energy} for misfit polarization along orthogonal directions in rectangular domains is uncoupled.  That is, the excess energy of for a general $\psi$ value can be rewritten using the solution for the $\psi=0$ elastic energy Eq.\ (\ref{eq:excess energy density}) as
            \begin{multline}
                \tilde{\cal{E}}_{\rm ex} ( \psi; w, h) = \sin^2 \psi ~ \tilde{\cal{E}}_{\rm ex} (\psi =0; h, w)\\
                +\cos^2 \psi ~ \tilde{\cal{E}}_{\rm ex} (\psi =0; w, h) .
                \label{eq: exmixed}
            \end{multline}
            Formally, this result follows from the orthogonality of strain modes in the series solutions detailed in Appendix \ref{app:orthogonality}.  Intuitively, we can understand this as result of a simple symmetry argument that elastic energy must be invariant under the reflection of the misfit polarization or shape through either the horizontal or vertical axis since mirroring the WJ particle shapes and domains will not change the (ground state) energy of the assemblies.  This argument implies that there can be no terms in the elastic energy proportional to the product of column and row curvatures $\kappa_1\kappa_2$, and that the only terms allowed at quadratic order are proportional to $\kappa_1^2$ and $\kappa_2^2$, resulting in the form of Eq.\ (\ref{eq: exmixed}).  
            
            Taken together, these arguments imply that both the ground state shapes {\it and} elastic energy behave as simple mixtures of the ``vertical'' and ``horizontal'' solutions when misfit polarization is not aligned to 2D lattice directions of the WJ reference domain.  Hence, this implies that the excess energy itself becomes symmetric with respect to exchange of wide and tall directions (i.e.under $(w,h) \to (h,w)$ for $\psi \to \pi/4$ where $\kappa_1 = \kappa_2$).  It is most useful to consider the implications of this superposition in terms of the infinite ribbon domains given by
            \begin{align}
                \tilde{\mathcal{E}}_{\rm ex}^{(\rm tall)}(\psi,w)&=
                \sin^2\psi\frac{w^2}{3+w^2} +\cos^2\psi\left(1-\frac{\tanh w}{w}\right)\label{eq:psiribbon tall}\\
                \tilde{\mathcal{E}}_{\rm ex}^{(\rm wide)}(\psi,h)&= \cos^2\psi\frac{h^2}{3+h^2} +\sin^2\psi\left(1-\frac{\tanh h}{h}\right)\label{eq:psiribbon wide},
            \end{align}
            with corresponding mesoscopic curvatures
            \begin{align}
                \kappa_c^{(\rm tall)}(\psi,w)=\frac{\kappa_0 \sin \psi}{1+w^2/3}\label{eq:psiribbon curvature tall}\\
                \kappa_c^{(\rm wide)}(\psi,h)=\frac{\kappa_0 \cos \psi}{1+h^2/3}.
                \label{eq:psiribbon curvature wide}
            \end{align}
            These results show that the off-lattice misfit polarization leads to a mixing of the two distinct elastic modes for anisotropic $\psi =0$ domains: the softer shear-bend mode in tall and mesoscopically-straight domains and the relatively stiffer stretch-bend mode in wide and mesoscopically-curved domains.  As a consequence, for a fixed domain size, the ground-state energy and mesoscopically-curved shapes of tall and wide domains are most distinct when the polarization is aligned to one of the two binding directions, and are most similar and degenerate when the polarization is equally vertical and horizontal at $\psi=\pi/4$.

        \subsection{Off-lattice ribbon domains ($\phi \neq 0$)}
            The prior sections show that two competing mechanisms of anisotropy selection underlie self-limiting ribbons domains of the WJ assembly model.  On one hand, anisotropy of binding tends to favor domains that grow longest along the relatively high binding energy directions, a generic effect of anisotropic edge energies in 2D assembly.  On the other hand, elastic costs of frustration are generically lowest for domains that extend along the misfit polarization direction $\Nv$.  In Sec.\ \ref{sec:psizero}, we showed that for $\psi=0$ WJ particles the elastic bias for ``tall'' ribbons stabilized these ribbons over ``wide'' (annular) ribbons even for binding that is weakly favorable for the latter (i.e.\ $\sigma_y \lesssim \sigma_x$).  Here we consider and analyze the broader possibility that polar frustration may select ground state domains that extend along off-lattice directions, i.e.\ other than the $\ev_2$ ($\phi =0$)  or $\ev_1$ ($\phi =\pi/2$) binding directions.
    
\begin{figure*}
                \centering
                \includegraphics[width=0.9\textwidth]{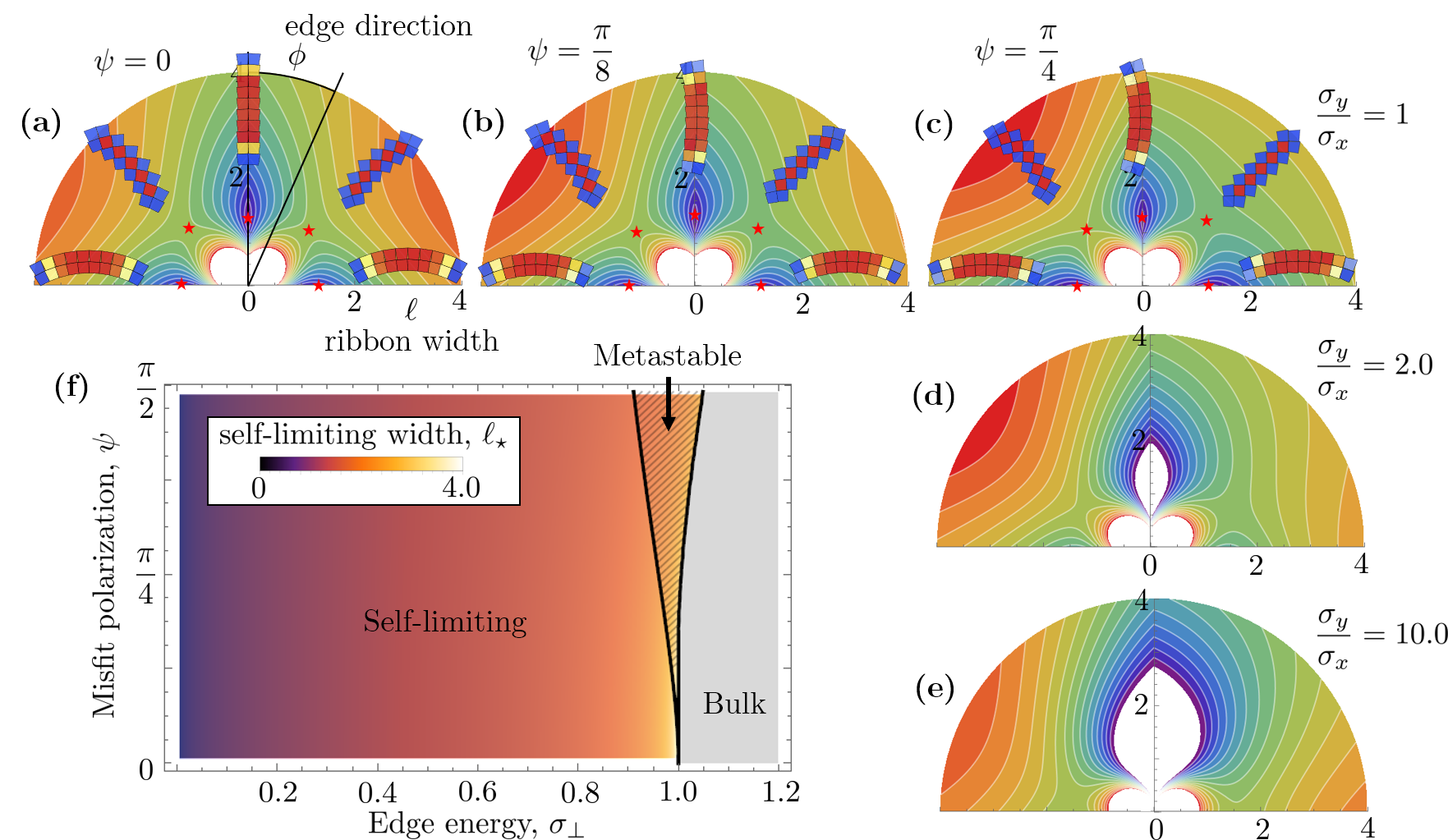}
                \caption{Energetics of off-lattice growth for ribbons with edges at an angle $\phi$ with respect to the vertical binding direction and thickness $\ell$.  \textbf{a,b,c)} Increasing misfit polarization angle $\psi=0$, $\pi/8$, and $\pi/8$.  Example assemblies are for the local minima $\phi=0$ and $\pm\pi/2$ and saddle point near $\phi=\pm\pi/4$.  \textbf{c,d,e)} Increasing ratio of binding energies with $\sqrt{\sigma_x^2+\sigma_y^2}=\sqrt{2}/2$.  \textbf{f)} Phase diagram of the transition between self-limiting and bulk growth for varying frustration polarization direction $\psi$ and edge cohesive strength $\sigma_{\perp}$.}
                \label{fig:binding direction}
            \end{figure*}

            To address this we consider the per-subunit assembly free energy of a ribbon whose long axis extends along a direction that makes angle $\phi$ relative to $\hat{\yv}$ in the 2D reference lattice (see Fig. \ref{fig:phi_definition}),
            \begin{multline}
                \frac{f(\psi,\phi;\ell)}{f_\infty}=\frac{\sigma_x|\cos\phi|+\sigma_y|\sin\phi|}{\ell}\\
                +\cos^2(\psi-\phi)\left(1-\frac{\tanh\ell}{\ell}\right)+\sin^2(\psi-\phi)\frac{\ell^2}{3+\ell^2},
             \end{multline}
             where $\ell$ characterizes the thickness of the narrow dimension of the ribbon (perpendicular to the long dimension).  For off-lattice directions $0<\phi<\pi/2$, the edge energy per unit length of the ribbon varies with $\phi$ according to the cohesive cost breaking additional bonds when cutting a lattice along directions that are not aligned to nearest-neighbor binding directions \cite{Rottman:1984}.  Based on this generalized ribbon model, we plot in Fig.\ \ref{fig:binding direction} the per-subunit free energy landscape as a function of finite-width $\ell$ and orientation $\phi$ for several cases of misfit polarization $\psi$ and binding anisotropy $\sigma_x/\sigma_y$.
    
            First, consider Figs.\ \ref{fig:binding direction}a-c which which show a sequence of energy landscapes for $\psi =0, \pi/8$, and $\pi/4$ for isotropic binding $\sigma_x=\sigma_y=0.5$.  All three cases show the existence of two local minima in the $(\ell, \phi)$ plane, corresponding to tall, vertically-oriented ($\phi =0$) and wide, horizontally-oriented ($\phi =\pm\pi/2$) ribbons with finite thickness.  Notably, even for $0<\psi<\pi/4$ these local minima do not shift away from the lattice directions, despite the fact that the elastic costs of frustration are lower (for a given thickness $\ell$) for ribbons extending along the polarization direction $\phi = \psi$.  Hence, all three landscapes are characterize by an intermediate unstable ribbon morphology at saddle points near $\phi=\pm\pi/4$, shown for each of the three cases.  
    
            In Figs.\ \ref{fig:binding direction}c-e, we consider the energy landscapes for fixed off-lattice polarization $\psi=\pi/4$ with increasing binding anisotropy, which show that the local minima corresponding to tall and wide ribbons can still exist, although for sufficient anisotropic binding the otherwise metastable wide ribbon morphology ultimately disappears, becoming thermodynamically unstable to bulk assembly in that direction.  
            
            Taken together, these assembly free energy landscapes demonstrate that changes in misfit polarization influence the ground state thermodynamics and gross morphological features of the self-limiting ribbon morphologies.  While ribbon morphologies that extend along the misfit polarization direction ($\phi=\psi)$ tend to have lower elastic costs, these results show that the possible orientations of equilibrium and metastable morphologies appear to be strongly locked to the underlying 2D lattice directions due the anisotropic biases of edge energy.  Mathematically, the relatively stronger effect of edge energy derives from its sharp and non-analytic dependence near to local minima (i.e.\ $|\phi - n \pi/2|$) in comparison to a much softer harmonic angular dependence (i.e.\ $(\phi - \psi)^2$) from elastic effects.  In Fig.\ \ref{fig:binding direction}, we plot the generalized equation of state $l_{\star}(\sigma_{\perp,\psi})$, the self-limiting ribbon width, where $\sigma_\perp$ is defined as the scaled edge energy for the free edge perpendicular to the ribbon's growth direction.  This shows that for any $\psi \neq 0$ or $\pi/4$ the transition between self-limiting and bulk states becomes at least weakly first order at critical cohesive strength that decreases continuously from $\sigma_c =1$ at $\psi= 0$ to a lower value at $\psi= \pi/4$.

	\section{Discussion and conclusions}

        In this article, we studied the ground-state thermodynamics of the 2D planar frustrated self-assembly of polar misfitting warped-jigsaw particles via simulations of discrete particle arrays and continuum elasticity theory.  Our results illuminate how anisotropic mesoscopic morphologies in the form of self-limiting multi-particle domains are selected by the distinct anisotropic features of the WJ particles themselves.  Much like with standard unfrustrated assembly systems, such as the Wulff construction of 2D crystals \cite{Rottman:1984}, we find that anisotropic inter-particle binding biases assemblies to grow relatively larger in directions corresponding to low edge energies.  However, in this case, we showed that {\it polarity} of the misfit direction is also transmitted to the multi-particle scale, so that even when binding energies are equal along orthogona directions, ground state shapes and thermodynamics of ribbons that grow ``tall" versus ``wide'' are mesoscopically distinguishable, each characterized by distinct mechanical modes of stress accumulation with the finite thickness of the domain.
    
        WJ assembly thermodynamics are distinct from other classes misfit symmetry, such as scalar or nematic, in that the equilibrium self-limiting morphologies are imprinted with a distinctly polar character that arises due to the polarity in particle shape at the local scale.  For example, tall ribbons of $\psi=0$ WJ particles exhibit asymmetric shapes at their top and bottom ends, while wide ribbons exhibit mesoscopic ``upwards" or ``downwards" bending, which itself is sensitive to the ribbon thickness.  Mesoscopic polarity arises because the microscopic asymmetry of WJ misfits breaks the otherwise 2-fold symmetry of the 2D domain anisotropy, unlike scalar or nematic misfit symmetries which preserve the underlying 2-fold symmetry.  As a result of the incommensurate symmetries of polar misfit and the background 2D order of assembly, ground state thermodynamics of WJ assembly exhibits a rich polymorphism, with multiple mesoscopically distinct self-limiting morphologies.   Shapes and thermodynamics of WJ assembly are controlled not only by the ratios of cohesion to elastic cost, as in standard GFA models, but also by the relative orientation of misfit polarity and 2D assembly order.  
    
        The WJ model and the dependence of the mesoscopic morphologies on misfit polarity provide new design parameters for experimental efforts to realize and control self-limiting assembly via microscopic misfit particle features such as shape and interactions.  The necessary features to implement WJ assembly are directly implementable, for example, using DNA origami based particles \cite{Berengut:2020, Hayakawa2024, Lee2025} or even so-called magnetic hand-shake particles \cite{Niu2019, Du2022}.  Both programmable subunit platforms combine the necessary ingredients to engineer and control the 2D misfitting shapes (e.g.\ trapezoids) of nanoscale subunits as well as their specific interactions needed to promote oriented edge bending, whose relative strengths could be dynamically tuned to stimulate transitions between the mesoscopically distinct self-limiting polymorphs.  
    
        \begin{figure}
            \centering
            \includegraphics[width=\columnwidth]{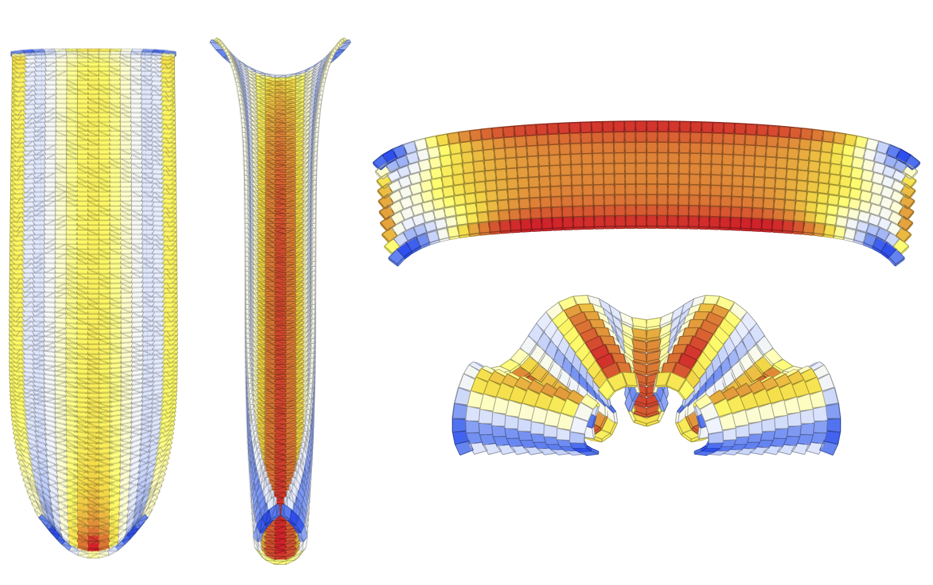}
            \caption{Examples of out-of-plane deformations of tall and wide assemblies when the out-of-plane bending modulus, related to subunit thickness, is decreased.}
            \label{fig:out of plane}
        \end{figure}
    
        Of course, the present theoretical study is limited in several respects that may be key to connecting with specific experimental systems.  Here we note two primary limitations which require further study.  Foremost, our study is restricted to ground states, essentially $T=0$ thermodynamics, whereas equilibrium assembly necessarily requires finite-temperature effects as well as finite range interations between subunits.  Notably, the equilibrium self-limitation behavior is predicted to depend on the ratio of inter-particle cohesion to inter-particle elastic stiffness, both features of which are mediated by the interactions that bind neighbor particles together.  As argued previously for simple models of inter-subunit binding \cite{Spivack:2022, Hall:2023}, equilibrium self-limitation places restrictions on the possible range of interactions.  For example, the thermodynamics of tall ribbons with self-limiting widths that are multiple WJ particles wide requires $k_x /k_y >1$, which can be achieved by introducing relatively short range binding in the $\ev_1$ direction to make rows stiffer to bend while introducing longer range binding in the $\ev_2$ direction to allow for softer shearing between rows.  How much the equilibrium windows of tall ribbon assembly can be extended by controlling interactions with anisotropic range, not to mention the potential kinetic limitations of finite-range interactions on reaching those equilibrium states, remains to be studied.  A second major open issue is the stability of the 2D planar domains to out-of-plane deformations and what kinds of complex 3D morphologies emerge.  As an example, we show in Fig.\ \ref{fig:out of plane} a simple extension of our planar WJ model that allows out-of-plane bending at a finite bending stiffness.  Notably, we observe that both wide and tall ribbon domains exhibit complex 3D buckled shapes when the out-of-plane bending stiffness falls below an apparent threshold.  What determines the threshold stiffness to suppress out-of-plane buckling in WJ domains, how the microscopic features such as the shape and interactions of WJ particles affects the out-of-plane morphologies, and in particular how misfit frustration is relaxed by deforming into the third dimension remains to be explored.
        
        Notwithstanding these limitations on our current understanding of WJ assembly, this study points to a more general classification of GFAs in terms of the misfit symmetry of the subunits and its resulting effects at the mesoscale.  Such a framework may be useful for analyzing the ingredients and outcomes of GFA for a much broader class of misfitting particles.  For example, Refs.\ \cite{Lenz:2017,Leroy:2023}, through a study of misfitting elastic polygons, have noted that for multi-protein assembly some measure of shape misfit is more likely the rule rather than the exception.  It is reasonable to expect that misfitting shape in such models could be mapped onto a multi-pole expansion in the misfit symmetry, and further that mesoscopic assembly outcomes (e.g.\ self-limiting domain size and shape) may be systematically correlated to the corresponding moments of misfit symmetry at the particle scale.  The notion of misfit also points the possibility of a generic, field-theoretic (e.g.\ Landau theory) approach to classifying and studying the interplay between inter-particle order and distinct symmetries of frustrated units.  Notably, recent progress along these lines has been made in the context of nematic liquid crystalline phases, which can be frustrated by a combination of distinct ``modes'' \cite{Selinger2022}: splay (polar), bend (polar), twist (psuedoscalar), and bi-axial splay (tensorial).  The general consequences for these distinct modes for finite-size domain selection in 3D nematically ordered assemblies remains to be understood \cite{Pollard_2021,daSilva_2021}, notwithstanding the much broader possibilities offered by distinct phases of long-range order (e.g.\ nematic, columnar, crystalline) in the assembled state.

    \section*{Acknowledgments}
        The authors are grateful to R. Matthew, I. Spivack, N. Hackney, J. Berengut, and L. Lee for valuable discussions about this work.  This study was supported by US National Science Foundation through awards NSF DMR-2028885, DMR-2349818 and the Brandeis Center for Bioinspired Soft Materials, an NSF MRSEC, DMR-2011846.  Simulation studies of the monomer model were performed on the UMass UNITY Cluster at the  Massachusetts Green High Performance Computing Center.
    
    \appendix
    \section{\label{app:sim details}Simulation details and energy relaxation}
        We model our discrete WJ particles as rigid bodies composed of pairs of elastic bonding sites on each of their four edges.  Let the position and frame of the particle be described by $\mathbf{R}$ and $\{\ev_1,\ev_2\}$, respectively, as shown in Fig.\ \ref{fig:model}c.  The attractive sites of the top, bottom, left, and right edges of the quadrilateral-shaped particle are respectively located at
        \begin{subequations}
            \begin{align}
                \mathbf{s}_{\textrm{top},\pm}&=\mathbf{R}\pm\frac{v}{2}\ev_1+\left(\frac{a}{2}\pm\frac{v}{2}\tan\alpha_2\right)\ev_2\\
                \mathbf{s}_{\textrm{bottom},\pm}&=\mathbf{R}\pm\frac{v}{2}\ev_1-\left(\frac{a}{2}\pm\frac{v}{2}\tan\alpha_2\right)\ev_2\\
                \mathbf{s}_{\textrm{left},\pm}&=\mathbf{R}-\left(\frac{a}{2}\pm\frac{v}{2}\tan\alpha_1\right)\ev_1\pm\frac{v}{2}\ev_2\\
                \mathbf{s}_{\textrm{right},\pm}&=\mathbf{R}+\left(\frac{a}{2}\pm\frac{v}{2}\tan\alpha_1\right)\ev_1\pm\frac{v}{2}\ev_2
            \end{align}
            \label{eq:attractive sites}%
        \end{subequations}
        where $\alpha_1$ and $\alpha_2$ are the taper angles of the left-right ($\ev_1$) and top-bottom ($\ev_2$) edges, respectively.  The $\pm$ represents the two interaction sites on each edge.  To simulate sheets of such particles, we arrange them on a reference square lattice at positions $\mathbf{R}^{(m,n)}=ma\hat{\xv}+na\hat{\yv}$ with the particle frames aligned with the lattice directions $\{\ev_1,\ev_2\}=\{\hat{\xv},\hat{\yv}\}$, and bind the specific binding sites together ($\mathbf{s}_{\rm bottom,\pm}^{(m,n+1)}$ with $\mathbf{s}_{\rm top,\pm}^{(m,n)}$ and $\mathbf{s}_{\rm left,\pm}^{(m+1,n)}$ with $\mathbf{s}_{\rm right,\pm}^{(m,n)}$) with zero rest length Hookean springs with stiffnesses $k_1$ and $k_2$ between the left-right and top-bottom edges of neighboring particles, respectively.

        Once a pre-assembled sheet of WJ particles with a given connectivity is created, we can relax the structure by computing the forces and torques acting on each particle and translating and rotating them according to overdamped equations of motion.  To be more precise, the position $\mathbf{R}$ of a particle is updated via
        \begin{eqnarray}
            \mathbf{R}(t+\delta t)=\mathbf{R}(t)+\delta t\mathbf{f}_{\rm total}
        \end{eqnarray}
        where $\mathbf{f}_{\rm total}=\sum_{\rm sites}\mathbf{f}_{\rm site}$ is the sum of the forces acting of the elastic binding sites of the particle and $\delta t$ is the step size.  The orientation $\theta$ of the particle, defined by $\cos\theta=\ev_2\cdot\hat{\yv}$, is updated via
        \begin{eqnarray}
            \theta(t+\delta t)=\theta(t)+\delta t|\boldsymbol{\tau}_{\rm total}|
        \end{eqnarray}
        where $\boldsymbol{\tau}_{\rm total}=\sum_{\rm sites}\mathbf{r}_{\rm site}\times\mathbf{f}_{\rm site}$ is the sum of torques due to the forces on the elastic binding sites and $\mathbf{r}_{\rm site}$ is the position of an elastic binding site relative to the center of the particle.  
    
    \section{\label{app:continuum}Continuum elasticity of WJ particle sheets}
        We here derive the continuum theory for describing $M\times N$ sheets of WJ particles.  As described in Appendix \ref{app:sim details}, we treat the WJ particles as rigid subunit with a pair of elastic binding sites on each edge.  We start with an $M\times N$ sheet arranged on a reference square lattice such that particle $(m,n)$ is located at $\mathbf{R}^{(m,n)}=ma\hat{\xv}+na\hat{\yv}+\mathbf{u}^{(m,n)}$, where $\mathbf{u}^{(m,n)}$ is a displacement relative to the reference configuration.  Using the positions of the elastic binding sites (Eqs.\ (\ref{eq:attractive sites}a-d)), we can compute the separations between the specific interaction sites between the top-bottom edges and left-right edges of neighboring subunits, which are given by
        \begin{subequations}
            \begin{multline}
                \Delta\mathbf{s}_{2,\pm}^{(m,n)}=\mathbf{s}_{\textrm{bottom},\pm}^{(m,n+1)}-\mathbf{s}_{\textrm{top},\pm}^{(m,n)}=a\hat{\yv}+\Delta_n\uv^{(m,n)}\\
                \pm\frac{v}{2}\Delta_n\ev_1^{(m,n)}-\left(a\pm v\tan\alpha_2\right)\langle\ev_2^{(m,n)}\rangle_n
            \end{multline}
            \begin{multline}
                \Delta\mathbf{s}_{1,\pm}^{(m,n)}=\mathbf{s}_{\textrm{right},\pm}^{(m+1,n)}-\mathbf{s}_{\textrm{left},\pm}^{(m,n)}=a\hat{\xv}+\Delta_m\uv^{(m,n)}\\
                -\left(a\pm v\tan\alpha_1\right)\langle\ev_1^{(m,n)}\rangle_m\pm\frac{v}{2}\Delta_m\ev_2^{(m,n)}
            \end{multline}
        \end{subequations}
        where we have defined $\Delta_n\uv^{(m,n)}=\uv^{(m,n+1)}-\uv^{(m,n)}$, $\Delta_n\ev_1^{(m,n)}=\ev_1^{(m,n+1)}-\ev_1^{(m,n)}$, and $\langle\ev_2^{(m,n)}\rangle_n=(\ev_2^{(m,n+1)}+\ev_2^{(m,n)})/2$ (and similarly for $\Delta_m$ and $\langle\rangle_m$ for the $(m+1,n)$ and $(m,n)$ subunits).  Assuming the interaction sites interact via Hookean springs with stiffnesses $k_1$ and $k_2$ for the left-right and top-bottom edges respectively, the energy of a sheet is
        \begin{align}
            \begin{split}
                E={}&\sum_{m,n}\frac{1}{2}k_1\left[\left(\Delta\mathbf{s}_{1,+}^{(m,n)}\right)^2+\left(\Delta\mathbf{s}_{1,-}^{(m,n)}\right)^2\right]\\
                {}&+\sum_{m,n}\frac{1}{2}k_2\left[\left(\Delta\mathbf{s}_{2,+}^{(m,n)}\right)^2+\left(\Delta\mathbf{s}_{2,-}^{(m,n)}\right)^2\right]\\
                ={}&\sum_{m,n}\frac{1}{2}k_1\left[\left(a\hat{\xv}+\Delta_m\uv^{(m,n)}-a\langle\ev_1^{(m,n)}\rangle_m\right)^2\right.\\
                {}&\left.+\frac{v^2}{4}\left(\Delta_m\ev_2^{(m,n)}-2\tan\alpha_1\langle\ev_1^{(m,n)}\rangle_m\right)^2\right]\\
                {}&+\sum_{m,n}\frac{1}{2}k_2\left[\left(a\hat{\yv}+\Delta_n\uv^{(m,n)}-a\langle\ev_2^{(m,n)}\rangle_n\right)^2\right.\\
                {}&\left.+\frac{v^2}{4}\left(\Delta_n\ev_1^{(m,n)}-2\tan\alpha_2\langle\ev_2^{(m,n)}\rangle_n\right)^2\right]
            \end{split}
        \end{align}
        The continuum limit can be obtained by taking $\Delta_m/a\rightarrow\partial_x$, $\Delta_n/a\rightarrow\partial_y$, and $\sum_{m,n}a^2\rightarrow\int dA$, resulting in
        \begin{multline}
            E=\int dA\left[\frac{1}{2}Y_1\left(\partial_x\uv+\hat{\xv}-\ev_1\right)^2+\frac{1}{2}B_1\left(\partial_x\ev_2-\kappa_1\ev_1\right)^2\right.\\
            +\left.\frac{1}{2}Y_2\left(\partial_y\uv+\hat{\yv}-\ev_2\right)^2+\frac{1}{2}B_2\left(\partial_y\ev_1-\kappa_2\ev_2\right)^2\right]
        \end{multline}
        where we identify the moduli and curvatures as
        \begin{equation}
            Y_{\alpha}=2k_{\alpha}, B_{\alpha}=\frac{1}{4}k_{\alpha}v^2
        \end{equation}
        and $\kappa_1=(2\tan\alpha_1)/a$ and $\kappa_2=(2\tan\alpha_2)/a$.  For small rotations $\theta$ from the reference state, we may write $\ev_1=\cos\theta\hat{\xv}+\sin\theta\hat{\yv}\simeq\hat{\xv}+\theta\hat{\yv}$ and $\ev_2=-\sin\theta\hat{\xv}+\cos\theta\hat{\yv}\simeq-\theta\hat{\xv}+\hat{\yv}$.  The linearized energy functional becomes
        \begin{multline}
            E=\int dA\left[\frac{1}{2}Y_x\left(\partial_x\uv-\theta\hat{\yv}\right)^2+\frac{1}{2}B_x\left(\partial_x\theta+\kappa_x\right)^2\right.\\
            +\left.\frac{1}{2}Y_y\left(\partial_y\uv+\theta\hat{\xv}\right)^2+\frac{1}{2}B_y\left(\partial_y\theta-\kappa_y\right)^2\right]
            \label{eqapp:elastic energy}
        \end{multline}
        Minimizing with respect to the displacement $\uv$ and rotations $\theta$, we obtain the Euler-Lagrange equations for force balance
        \begin{subequations}
            \begin{align}
                Y_x\partial_x^2u_x+Y_y\partial_y^2u_x&=-\partial_y\theta\label{eqapp:ux force}\\
                Y_x\partial_x^2u_y+Y_y\partial_y^2u_y&=\partial_x\theta\label{eqapp:uy force}
            \end{align}
        \end{subequations}
        and torque balance
        \begin{equation}
            B_x \partial_x^2 \theta +B_y \partial_y^2 \theta =  -Y_{x} (\partial_x u_y - \theta) + Y_{y} (\partial_y u_x + \theta)\label{eqapp:theta torque}
        \end{equation}
        The boundary conditions for the free edges along $x=\pm W/2$ are
        \begin{subequations}
            \begin{align}
                \partial_xu_x|_{x=\pm W/2}&=0\label{eqapp:ux x edge}\\
                (\partial_xu_y-\theta)|_{x=\pm W/2}&=0\label{eqapp:uy x edge}\\
                (\partial_x\theta+\kappa_x)|_{x=\pm W/2}&=0\label{eqapp:theta x edge}
            \end{align}
        \end{subequations}
        and along $y=\pm H/2$ are
        \begin{subequations}
            \begin{align}
                (\partial_yu_x+\theta)|_{y\pm H/2}&=0\label{eqapp:ux y edge}\\
                \partial_yu_y|_{y=\pm H/2}&=0\label{eqapp:uy y edge}\\
                (\partial_y\theta-\kappa_y)|_{y\pm H/2}&=0\label{eqapp:theta y edge}
            \end{align}
        \end{subequations}

    \section{\label{app:solution}Analytical solution and approximation}
        Note that due to linearity, we can solve the continuum theory by splitting it into two separate problems: one for $\kappa_x\ne0, \kappa_y=0$ and the other for $\kappa_x=0, \kappa_y\ne0$.  We focus on $\kappa_x\ne0,\kappa_y=0$ since the other case can simply be obtained by swapping indices $x\leftrightarrow y$ and curvatures $\kappa_x\leftrightarrow-\kappa_y$.  

        To start, it will be useful to briefly summarize the solution for infinitely tall ribbons with $\kappa_x\ne0, \kappa_y=0$ in \cite{Spivack:2022}, as we will use it to construct the general solution.  Consider such a infinite ribbon bound between $-W/2\le x\le W/2$.  Since there will be no $y$ dependence due to symmetry, the Euler-Lagrange equations become
        \begin{subequations}
            \begin{align}
                Y_x\partial_x^2u_x&=0\\
                Y_x\partial_x^2u_y&=\partial_x\theta\\
                B_x\partial_x^2\theta&=-Y_x\partial_xu_y+(Y_x+Y_y)\theta
            \end{align}
        \end{subequations}
        The boundary conditions along $x=\pm W/2$ are the same as Eqs.\ (\ref{eqapp:ux x edge}), (\ref{eqapp:uy x edge}), and (\ref{eqapp:theta x edge}).  This system of equations can easily be solved for the displacement and rotation fields, which are
        \begin{subequations}
            \begin{align}
                u_x^{(\infty)}(x)&=0\\
                u_y^{(\infty)}(x)&=-\frac{\kappa_xB_x}{Y_y}\frac{\cosh\frac{x}{\sqrt{B_x/Y_y}}}{\cosh\frac{W}{2\sqrt{B_x/Y_y}}}\\
                \theta^{(\infty)}(x)&=-\kappa_x\sqrt{\frac{B_x}{Y_y}}\frac{\sinh\frac{x}{\sqrt{B_x/Y_y}}}{\cosh\frac{W}{2\sqrt{B_x/Y_y}}}
            \end{align}
            \label{eq:infinite tall solutions}%
        \end{subequations}
        Using these results, we construct the solution for a finite domain bound by $-W/2\le x\le W/2$ and $-H/2\le y\le H/2$ as follows.  We write the displacement and rotation fields as
        \begin{subequations}
            \begin{align}
                u_x(x,y)&=u_x^{(\infty)}(x)+U_x(x,y)\\
                u_y(x,y)&=u_y^{(\infty)}(x)+U_y(x,y)\\
                \theta(x,y)&=\theta^{(\infty)}(x)+\Theta(x,y)
            \end{align}
        \end{subequations}
        Substituting these into Eqs.\ (\ref{eqapp:ux force}), (\ref{eqapp:uy force}), and (\ref{eqapp:theta torque}), we find that $U_x(x,y)$, $U_y(x,y)$, and $\Theta(x,y)$ satisfy the same force and torque balance equations
        \begin{subequations}
            \begin{align}
                Y_x\partial_x^2U_x+Y_y\partial_y^2U_x&=-\partial_y\Theta\label{eqapp:Ux force}\\
                Y_x\partial_x^2U_y+Y_y\partial_y^2U_y&=\partial_x\Theta\label{eqapp:Uy force}\\
                B_x\partial_x^2\Theta+B_y\partial_y^2\Theta&=-Y_x(\partial_xU_y-\Theta)+Y_y(\partial_yU_x+\Theta)\label{eqapp:Theta torque}
            \end{align}
        \end{subequations}
        with the boundary conditions
        \begin{subequations}
            \begin{align}
                \partial_xU_x|_{x=\pm W/2}&=0\label{eqapp:Ux x edge}\\
                (\partial_xU_y-\Theta)|_{x=\pm W/2}&=0\label{eqapp:Uy x edge}\\
                \partial_x\Theta|_{x=\pm W/2}&=0\label{eqapp:Theta x edge}
            \end{align}
        \end{subequations}
        \begin{subequations}
            \begin{align}
                (\partial_yU_x+\Theta)|_{y=\pm H/2}&=-\theta^{(\infty)}(x)\label{eqapp:Ux y edge}\\
                \partial_yU_y|_{y=\pm H/2}&=0\label{eqapp:Uy y edge}\\
                \partial_y\Theta|_{y=\pm H/2}&=0\label{eqapp:Theta y edge}
            \end{align}
        \end{subequations}
        We start with constructing $\Theta(x,y)$.  Note that due to symmetry, $\Theta(x,y)$ must be odd in $x$.  From the $\Theta$ boundary conditions (Eqs.\ (\ref{eqapp:Theta x edge}) and (\ref{eqapp:Theta y edge})), we can write
        \begin{multline}
            \Theta(x,y)=\sum_{\substack{m\textrm{ odd}\\n\textrm{ even}}}A_{m,n}\sin\frac{m\pi x}{W}\cos\frac{n\pi y}{H}\\
            +\sum_{\substack{m\textrm{ odd}\\n\textrm{ odd}}}A_{m,n}\sin\frac{m\pi x}{W}\sin\frac{n\pi y}{H}
        \end{multline}
        Similarly, $U_x(x,y)$ must also be odd in $x$.  From the boundary condition Eq.\ (\ref{eqapp:Ux x edge}), we can write
        \begin{equation}
            U_x(x,y)=\sum_{m\textrm{ odd}}\sin\frac{m\pi x}{W}F_m(y)
        \end{equation}
        for some function $F_m(y)$.  Substituting this into Eq.\ (\ref{eqapp:Ux force}) and using the orthogonality of the set $\left\{\sin\frac{m\pi x}{W}\right\}_{m\textrm{ odd}}$, we have that $F_m(y)$ satisfies
        \begin{multline}
            \frac{d^2F_m}{dy^2}-\frac{m^2\pi^2Y_x}{W^2Y_y}F_m\\
            =\sum_{n\textrm{ even}}\frac{n\pi}{H}A_{m,n}\sin\frac{n\pi y}{H}-\sum_{n\textrm{ odd}}\frac{n\pi}{H}A_{m,n}\cos\frac{n\pi y}{H}
        \end{multline}
        the general solution of which is
        \begin{multline}
            F_m(y)=c_m\cosh\frac{m\pi}{W}\sqrt{\frac{Y_x}{Y_y}}y+s_m\sinh\frac{m\pi}{W}\sqrt{\frac{Y_x}{Y_y}}y\\
            -\sum_{n\textrm{ even}}\frac{\frac{n\pi}{H}}{\frac{m^2\pi^2Y_x}{W^2Y_y}+\frac{n^2\pi^2}{H^2}}A_{m,n}\sin\frac{n\pi y}{H}\\
            +\sum_{n\textrm{ odd}}\frac{\frac{n\pi}{H}}{\frac{m^2\pi^2Y_x}{W^2Y_y}+\frac{n^2\pi^2}{H^2}}A_{m,n}\cos\frac{n\pi y}{H}
        \end{multline}
        The boundary condition Eq.\ (\ref{eqapp:Ux y edge}) tells us that
        \begin{multline}
                \left.\frac{dF_m}{dy}\right|_{y=\pm\frac{H}{2}}=\frac{4B_x\kappa_x\sin\frac{m\pi}{2}}{W\left(B_x\frac{m^2\pi^2}{W^2}+Y_y\right)}\\
                -\sum_{n\textrm{ even}}A_{m,n}\cos\frac{n\pi}{2}\mp\sum_{n\textrm{ odd}}A_{m,n}\sin\frac{n\pi}{2}
        \end{multline}
        which gives
        \begin{subequations}
            \begin{align}
                \begin{split}
                    s_m={}&\frac{1}{\frac{m\pi}{W}\sqrt{\frac{Y_x}{Y_y}}\cosh\frac{m\pi H}{2W}\sqrt{\frac{Y_x}{Y_y}}}\left[\frac{4B_x\kappa_x\sin\frac{m\pi}{2}}{W\left(B_x\frac{m^2\pi^2}{W^2}+Y_y\right)}\right.\\
                    &{}\left.-\sum_{n\textrm{ even}}\frac{\frac{m^2Y_x}{W^2Y_y}}{\frac{m^2Y_x}{W^2Y_y}+\frac{n^2}{H^2}}A_{m,n}\cos\frac{n\pi}{2}\right]
                    \label{eqapp:sm}
                \end{split}\\
                \begin{split}
                    c_m={}&-\frac{1}{\frac{m\pi}{W}\sqrt{\frac{Y_x}{Y_y}}\sinh\frac{m\pi H}{2W}\sqrt{\frac{Y_x}{Y_y}}}\\
                    &{}\times\sum_{n\textrm{ odd}}\frac{\frac{m^2Y_x}{W^2Y_y}}{\frac{m^2Y_x}{W^2Y_y}+\frac{n^2}{H^2}}A_{m,n}\sin\frac{n\pi}{2}
                    \label{eqapp:cm}
                \end{split}
            \end{align}
        \end{subequations}
        Finally, for $U_y(x,y)$, the boundary condition Eq.\ (\ref{eqapp:Uy y edge}) allows us to write
        \begin{equation}
            U_y(x,y)=\sum_{n\textrm{ even}}G_n(x)\cos\frac{n\pi y}{H}+\sum_{n\textrm{ odd}}G_n(x)\sin\frac{n\pi y}{H}
        \end{equation}
        Substituting into Eq.\ (\ref{eqapp:Uy force}), we find that $G_n(x)$ satisfies
        \begin{multline}
            \frac{d^2G_n}{dx^2}-\frac{n^2\pi^2Y_y}{H^2Y_x}G_n=\sum_{m\textrm{ odd}}\frac{m\pi}{W}A_{m,n}\cos\frac{m\pi x}{W}
        \end{multline}
        By symmetry, $U_y(x,y)$ must be even in $x$, and so the general solution for $G_n(x)$ can be written as
        \begin{multline}
            G_n(x)=g_n\cosh\frac{n\pi}{H}\sqrt{\frac{Y_y}{Y_x}}x\\
            -\sum_{m\textrm{ odd}}\frac{\frac{m\pi}{W}}{\frac{m^2\pi^2}{W^2}+\frac{n^2\pi^2Y_y}{H^2Y_x}}A_{m,n}\cos\frac{m\pi x}{W}
        \end{multline}
        The boundary condition Eq.\ (\ref{eqapp:Uy x edge}) tell us that
        \begin{equation}
            \left.\frac{dG_n}{dx}\right|_{x=\pm\frac{W}{2}}=\pm\sum_{m\textrm{ odd}}A_{m,n}\sin\frac{m\pi}{2}
        \end{equation}
        which gives for $n>0$
        \begin{multline}
            g_n=\frac{1}{\frac{n\pi}{H}\sqrt{\frac{Y_y}{Y_x}}\sinh\frac{n\pi W}{2H}\sqrt{\frac{Y_y}{Y_x}}}\\
            \times\sum_{m\textrm{ odd}}\frac{\frac{n^2Y_y}{H^2Y_x}}{\frac{m^2}{W^2}+\frac{n^2Y_y}{H^2Y_x}}A_{m,n}\sin\frac{m\pi}{2}
            \label{eqapp:gn}
        \end{multline}
        Note that $g_0$ just gives a global translation, so we pick $g_0=0$ for simplicity.  To determine an equation for $A_{m,n}$, we take the constructions for $\Theta$, $U_x$, and $U_y$ and substitute them into Eq.\ (\ref{eqapp:Theta torque}).  Using the orthogonality of the sines and cosines, we have for $n$ even
        \begin{widetext}
            \begin{multline}
                -\left(B_x\frac{m^2\pi^2}{W^2}+B_y\frac{n^2\pi^2}{H^2}\right)A_{m,n}=-Y_x\frac{\frac{4n^2}{WH^2}\cosh\frac{n\pi W}{2H}\sqrt{\frac{Y_y}{Y_x}}\sin\frac{m\pi}{2}}{\frac{m^2Y_x}{W^2Y_y}+\frac{n^2}{H^2}}g_n-Y_x\frac{\frac{m^2}{W^2}}{\frac{m^2}{W^2}+\frac{n^2Y_y}{H^2Y_x}}A_{m,n}\\
                +Y_y\frac{2(2-\delta_{n,0})\frac{m^2}{HW^2}\sinh\frac{m\pi H}{2W}\sqrt{\frac{Y_x}{Y_y}}\cos\frac{n\pi}{2}}{\frac{m^2}{W^2}+\frac{n^2Y_y}{H^2Y_x}}s_m-Y_y\frac{\frac{n^2}{H^2}}{\frac{m^2Y_x}{W^2Y_y}+\frac{n^2}{H^2}}A_{m,n}+(Y_x+Y_y)A_{m,n}
                \label{eqapp:Amn even}
            \end{multline}
            and for $n$ odd
            \begin{multline}
                -\left(B_x\frac{m^2\pi^2}{W^2}+B_y\frac{n^2\pi^2}{H^2}\right)A_{m,n}=-Y_x\frac{\frac{4n^2}{H^2}\cosh\frac{n\pi W}{2H}\sqrt{\frac{Y_y}{Y_x}}\sin\frac{m\pi}{2}}{\frac{m^2Y_x}{W^2Y_y}+\frac{n^2}{H^2}}g_n-Y_x\frac{\frac{m^2}{W^2}}{\frac{m^2}{W^2}+\frac{n^2Y_y}{H^2Y_x}}A_{m,n}\\
                +Y_y\frac{\frac{4m^2}{HW^2}\cosh\frac{m\pi H}{2W}\sqrt{\frac{Y_x}{Y_y}}\sin\frac{n\pi}{2}}{\frac{m^2}{W^2}+\frac{n^2Y_y}{H^2Y_x}}c_m-Y_y\frac{\frac{n^2}{H^2}}{\frac{m^2Y_x}{W^2Y_y}+\frac{n^2}{H^2}}A_{m,n}+(Y_x+Y_y)A_{m,n}
                \label{eqapp:Amn odd}
            \end{multline}
        \end{widetext}
        Note that for $n$ odd, $g_n$ and $c_m$ are sums of $A_{m,n}$.  By inspection of Eq.\ (\ref{eqapp:Amn odd}), we can conclude that $A_{m,n}=0$ for all odd $n$ satisfies Eq.\ (\ref{eqapp:Amn odd}).  Consequently, this means that $c_m=0$ for all $m$ and $g_n=0$ for odd $n$.  Thus, we find that $\Theta(x,y)$ and $U_y(x,y)$ are even in $y$ and $U_x(x,y)$ is odd in $y$.  We summarize the series solution.
        \begin{widetext}
            \begin{subequations}
                \begin{align}
                    \Theta(x,y)={}&\sum_{\substack{m\textrm{ odd}\\n\textrm{ even}}}A_{m,n}\sin\frac{m\pi x}{W}\cos\frac{n\pi y}{H}\\
                    U_x(x,y)={}&\sum_{m\textrm{ odd}}s_m\sin\frac{m\pi x}{W}\sinh\frac{m\pi}{W}\sqrt{\frac{Y_x}{Y_y}}y-\sum_{\substack{m\textrm{ odd}\\n\textrm{ even}}}\frac{\frac{n\pi}{H}}{\frac{m^2\pi^2Y_x}{W^2Y_y}+\frac{n^2\pi^2}{H^2}}A_{m,n}\sin\frac{m\pi x}{W}\sin\frac{n\pi y}{H}\\
                    U_y(x,y)={}&\sum_{n\textrm{ even}}g_n\cosh\frac{n\pi}{H}\sqrt{\frac{Y_y}{Y_x}}x\cos\frac{n\pi y}{H}-\sum_{\substack{m\textrm{ odd}\\n\textrm{ even}}}\frac{\frac{m\pi}{W}}{\frac{m^2\pi^2}{W^2}+\frac{n^2\pi^2Y_y}{H^2Y_x}}A_{m,n}\cos\frac{m\pi x}{W}\cos\frac{n\pi y}{H}
                \end{align}
                \label{eq:series solution psi 0}
            \end{subequations}
        \end{widetext}
        where $s_m$, $g_n$, and $A_{m,n}$ for odd $m$ and even $n$ are given by Eqs.\ (\ref{eqapp:sm}), (\ref{eqapp:gn}), and (\ref{eqapp:Amn even}), respectively.  Unfortunately, due to $s_m$ and $g_n$ depending on infinite sums of $A_{m,n}$, we effectively have an infinite system of equations for determining an infinite set of coefficients, which is impossible to solve.  However, in Appendix \ref{app:approximation}, we consider approximating the solution by truncating the series solution.

        So far, we solved the case of $\kappa_x\ne0, \kappa_y=0$.  Fortunately, we do not need to repeat the same cumbersome calculation for $\kappa_x=0,\kappa_y\ne0$.  All that needs to be done is to take the $\kappa_x\ne0,\kappa_y=0$ solution and simply swap all indices $x\leftrightarrow y$, lengths $W\leftrightarrow H$, and curvatures $\kappa_x\leftrightarrow-\kappa_y$, and sum up the corresponding displacement and rotation fields.  Putting this all together, we have that the general solution for arbitrary $\kappa_x,\kappa_y$ is
        \begin{widetext}
            \begin{subequations}
                \begin{align}
                    \begin{split}
                        \theta(x,y)={}&-\kappa_x\sqrt{\frac{B_x}{Y_y}}\frac{\sinh\frac{x}{\sqrt{B_x/Y_y}}}{\cosh\frac{W}{2\sqrt{B_x/Y_y}}}+\sum_{\substack{m\textrm{ odd}\\n\textrm{ even}}}A_{m,n}\sin\frac{m\pi x}{W}\cos\frac{n\pi y}{H}\\
                        &{}+\kappa_y\sqrt{\frac{B_y}{Y_x}}\frac{\sinh\frac{y}{\sqrt{B_y/Y_x}}}{\cosh\frac{H}{2\sqrt{B_y/Y_x}}}+\sum_{\substack{m\textrm{ odd}\\n\textrm{ even}}}A_{m,n}'\sin\frac{m\pi y}{H}\cos\frac{n\pi x}{W}
                        \label{eqapp:theta general}
                    \end{split}\\
                    \begin{split}
                        u_x(x,y)={}&\sum_{m\textrm{ odd}}s_m\sin\frac{m\pi x}{W}\sinh\frac{m\pi}{W}\sqrt{\frac{Y_x}{Y_y}}y-\sum_{\substack{m\textrm{ odd}\\n\textrm{ even}}}\frac{\frac{n\pi}{H}}{\frac{m^2\pi^2Y_x}{W^2Y_y}+\frac{n^2\pi^2}{H^2}}A_{m,n}\sin\frac{m\pi x}{W}\sin\frac{n\pi y}{H}\\
                        &{}+\frac{\kappa_yB_y}{Y_x}\frac{\cosh\frac{y}{\sqrt{B_y/Y_x}}}{\cosh\frac{H}{2\sqrt{B_y/Y_x}}}+\sum_{n\textrm{ even}}g_n'\cosh\frac{n\pi}{W}\sqrt{\frac{Y_x}{Y_y}}y\cos\frac{n\pi x}{W}-\sum_{\substack{m\textrm{ odd}\\n\textrm{ even}}}\frac{\frac{m\pi}{H}}{\frac{m^2\pi^2}{H^2}+\frac{n^2\pi^2Y_x}{W^2Y_y}}A_{m,n}'\cos\frac{m\pi y}{H}\cos\frac{n\pi x}{W}
                        \label{eqapp:ux general}
                    \end{split}\\
                    \begin{split}
                        u_y(x,y)={}&-\frac{\kappa_xB_x}{Y_y}\frac{\cosh\frac{x}{\sqrt{B_x/Y_y}}}{\cosh\frac{W}{2\sqrt{B_x/Y_y}}}+\sum_{n\textrm{ even}}g_n\cosh\frac{n\pi}{H}\sqrt{\frac{Y_y}{Y_x}}x\cos\frac{n\pi y}{H}-\sum_{\substack{m\textrm{ odd}\\n\textrm{ even}}}\frac{\frac{m\pi}{W}}{\frac{m^2\pi^2}{W^2}+\frac{n^2\pi^2Y_y}{H^2Y_x}}A_{m,n}\cos\frac{m\pi x}{W}\cos\frac{n\pi y}{H}\\
                        &{}+\sum_{m\textrm{ odd}}s_m'\sin\frac{m\pi y}{H}\sinh\frac{m\pi}{H}\sqrt{\frac{Y_y}{Y_x}}x-\sum_{\substack{m\textrm{ odd}\\n\textrm{ even}}}\frac{\frac{n\pi}{W}}{\frac{m^2\pi^2Y_y}{H^2Y_x}+\frac{n^2\pi^2}{W^2}}A_{m,n}'\sin\frac{m\pi y}{H}\sin\frac{n\pi x}{W}
                        \label{eqapp:uy general}
                    \end{split}
                \end{align}
            \end{subequations}
        \end{widetext}
        where $s_m'$, $g_n'$, and $A_{m,n}'$ satisfy similar relations as Eqs.\ (\ref{eqapp:sm}), (\ref{eqapp:gn}), and (\ref{eqapp:Amn even}) but with the swaps $x\leftrightarrow y$, $W\leftrightarrow H$, and $\kappa_x\leftrightarrow-\kappa_y$.  Note that these solutions are written so that the terms in the first line are proportional to $\kappa_x$ while those in the second line are proportional to $\kappa_y$.

        \subsection{\label{app:orthogonality}Absence of $\kappa_x\kappa_y$ terms in the elastic energy}
            Naturally, we are interested in the elastic energy for general $\kappa_x,\kappa_y$, which should contain terms proportional to $\kappa_x^2$, $\kappa_x\kappa_y$, and $\kappa_y^2$.  We show that the elastic energy contains no $\kappa_x\kappa_y$ terms i.e.\ the elastic energy can be written as the sum of the individual elastic energies for $\kappa_x\ne0,\kappa_y=0$ and $\kappa_x=0,\kappa_y\ne0$.  To see this, let's write the solutions as
            \begin{subequations}
                \begin{align}
                    \theta(x,y;\kappa_x,\kappa_y)={}&\theta(x,y;\kappa_x,0)+\theta(x,y;0,\kappa_y)\\
                    u_x(x,y;\kappa_x,\kappa_y)={}&u_x(x,y;\kappa_x,0)+u_x(x,y;0,\kappa_y)\\
                    u_y(x,y;\kappa_x,\kappa_y)={}&u_y(x,y;\kappa_x,0)+u_y(x,y;0,\kappa_y)
                \end{align}
            \end{subequations}
            The parity of each piece is as follows: $\theta(x,y;\kappa_x,0)$ is odd in $x$ and even in $y$ while $\theta(x,y;0,\kappa_y)$ is even in $x$ and odd in $y$; $u_x(x,y;\kappa_x,0)$ is odd in both $x$ and $y$ while $u_x(x,y;0,\kappa_y)$ is even in both $x$ and $y$; and finally $u_y(x,y;\kappa_x,0)$ is even in both $x$ and $y$ while $u_y(x,y;0,\kappa_y)$ is odd in both $x$ and $y$.  
            
            Let's examine each term in the elastic energy Eq.\ (\ref{eqapp:elastic energy}). We suppress the $x,y$ coordinates to keep things simple.  The bending energy terms are
            \begin{align}
                \begin{split}
                    (\partial_x\theta+\kappa_x)^2={}&[\partial_x\theta(\kappa_x,0)+\kappa_x]^2+[\partial_x\theta(0,\kappa_y)]^2\\
                    &{}+2[\partial_x\theta(\kappa_x,0)+\kappa_x]\partial_x\theta(0,\kappa_y)
                \end{split}\\
                \begin{split}
                    (\partial_y\theta-\kappa_y)^2={}&[\partial_y\theta(\kappa_x,0)]^2+[\partial_y\theta(0,\kappa_y)-\kappa_y]^2\\
                    &{}+2\partial_y\theta(\kappa_x,0)[\partial_y\theta(0,\kappa_y)-\kappa_y]
                \end{split}
            \end{align}
            Since $\partial_x\theta(\kappa_x,0)+\kappa_x$ is even in both $x$ and $y$ while $\partial_x\theta(0,\kappa_y)$ is odd in both $x$ and $y$, their product integrates to zero.  Similarly, since $\partial_y\theta(\kappa_x,0)$ is odd in both $x$ and $y$ while $\partial_y\theta(0,\kappa_y)-\kappa_y$ is even in both $x$ and $y$, their product also integrates to zero.  For the stretching energy terms, we have
            \begin{multline}
                (\partial_xu_x)^2=[\partial_xu_x(\kappa_x,0)]^2+[\partial_xu_x(0,\kappa_y)]^2\\
                +2\partial_xu_x(\kappa_x,0)\partial_xu_x(0,\kappa_y)
            \end{multline}
            \begin{multline}
                (\partial_xu_y-\theta)^2\\
                =[\partial_xu_y(\kappa_x,0)-\theta(\kappa_x,0)]^2+[\partial_xu_y(0,\kappa_y)-\theta(0,\kappa_y)]^2\\
                +2[\partial_xu_y(\kappa_x,0)-\theta(\kappa_x,0)][\partial_xu_y(0,\kappa_y)-\theta(0,\kappa_y)]
            \end{multline}
            \begin{multline}
                (\partial_yu_x+\theta)^2\\
                =[\partial_yu_x(\kappa_x,0)+\theta(\kappa_x,0)]^2
                +[\partial_yu_x(0,\kappa_y)+\theta(0,\kappa_y)]^2\\
                +2[\partial_yu_x(\kappa_x,0)+\theta(\kappa_x,0)][\partial_yu_x(0,\kappa_y)+\theta(0,\kappa_y)]
            \end{multline}
            \begin{multline}
                (\partial_yu_y)^2=[\partial_yu_y(\kappa_x,0)]^2+[\partial_yu_y(0,\kappa_y)]^2\\
                +2\partial_yu_y(\kappa_x,0)\partial_yu_y(0,\kappa_y)
            \end{multline}
            Following the same procedure, note that $\partial_xu_x(\kappa_x,0)$ is even in $x$ and odd in $y$ while $\partial_xu_x(0,\kappa_y)$ is odd in $x$ and even in $y$; $\partial_xu_y(\kappa_x,0)-\theta(\kappa_x,0)$ is odd in $x$ and even in $y$ while $\partial_xu_y(0,\kappa_y)-\theta(0,\kappa_y)$ is even in $x$ and odd in $y$; $\partial_yu_x(\kappa_x,0)+\theta(\kappa_x,0)$ is odd in $x$ and even in $y$ while $\partial_yu_x(0,\kappa_y)+\theta(0,\kappa_y)$ is even in $x$ and odd in $y$; and finally $\partial_yu_y(\kappa_x,0)$ is even in $x$ and odd in $y$ while $\partial_yu_y(0,\kappa_y)$ is odd in $x$ and even in $y$.  Therefore, the products of these terms integrate to zero, and so we see that the elastic energy can be decomposed into the sum of the elastic energies for $\kappa_x\ne0,\kappa_y=0$ and $\kappa_x=0,\kappa_y\ne0$, which are proportional to $\kappa_x^2$ and $\kappa_y^2$, respectively.

            Note that this result could also have been obtained from a simple symmetry argument.  Flipping the sign of $\kappa_x$ or $\kappa_y$ simply creates a subunit that is reflected across $\ev_1$ or $\ev_2$, respectively.  All this does is create an assembly that is a reflection of the original assembly, which means that the energy must remain the same.  Therefore, there cannot be any terms proportional to $\kappa_x\kappa_y$ in the elastic energy.

        \subsection{\label{app:approximation}Truncated series approximation}
            The solution of the coupled PDEs Eqs.\ (\ref{eq:force balance}) and (\ref{eq:torque balance}) requires solving an infinite system of equations Eq.\ (\ref{eqapp:Amn even}) for all the coefficients $A_{m,n}$ for all $n$ even, which therefore cannot be done in closed-form.
            
            To acquire a closed-form expression for the elastic ground states, we perform a truncation on the series solution Eq.\ (\ref{eq:series solution psi 0}).  We here focus on the case of $\kappa_x\ne0,\kappa_y=0$ to keep things simple since the general solution can be obtained by simply swapping indices as explained in Appendix \ref{app:solution}.  To determine how to truncate the solution, we make the following observation on the behavior of $u_x$ and $u_y$when going from an infinitely tall assembly $N\rightarrow\infty$ to one that is finite in height.  Note that for the infinite tall structures (Eqs.\ (\ref{eq:infinite tall solutions}a-c)), the row extension or compression is uniform all along $y$ with $u_x=0$ and vertical displacements have an exponential-like profile $u_y\sim\cosh(x/\sqrt{B_x/Y_y})$.  When $N$ is finite, two things occur:
            \begin{itemize}
                \item The extension and compression is no longer uniform all along $y$.  Instead, as shown in Fig.\ \ref{fig:elastic modes combined}, the top edge of an assembly experiences extension while the bottom edge, compression.  In other words, $u_x$ must have some nonzero $y$ dependence, particularly concentrated near the top and bottom edges of the assembly.

                \item The vertical displacements $u_y$ switch from a $\cosh$ profile where there can be a flattened core to a profile that resembles a circular arc throughout most of each row, as shown in Fig.\ \ref{fig:excess energy}d.  In other words, the largest correction to $u_y$ will be the change in its $x$ dependence when the height is finite.  Any variation along $y$ is essentially a higher order i.e.\ the top rows curve similarly to the bottom rows.
            \end{itemize}
            These physical observations informs us about an approximation.  We mostly need to account for the $x$ dependence changes of $u_y$ (from $\cosh$ to parabolic or circular) and the leading order dependence of $u_x$ on $y$ since $u_x=0$ in the infinitely tall ribbons.  This can be achieved by taking the leading $n$ mode i.e.\ $n=0$ and setting $A_{m,n}=0$ for $n>0$.  The result is the following approximation
            \begin{subequations}
                \begin{align}
                    \theta(x,y)&=-\kappa_x\sqrt{\frac{B_x}{Y_y}}\frac{\sinh\frac{x}{\sqrt{B_x/Y_y}}}{\cosh\frac{W}{2\sqrt{B_x/Y_y}}}+\sum_{m\textrm{ odd}}A_m\sin\frac{m\pi x}{W}\\
                    u_x(x,y)&=\sum_{m\textrm{ odd}}s_m\sin\frac{m\pi x}{W}\sinh\frac{m\pi}{W}\sqrt{\frac{Y_x}{Y_y}}y\\
                    u_y(x,y)&=-\frac{\kappa_xB_x}{Y_y}\frac{\cosh\frac{x}{\sqrt{B_x/Y_y}}}{\cosh\frac{W}{2\sqrt{B_x/Y_y}}}+\sum_{m\textrm{ odd}}\frac{W}{m\pi}A_m\cos\frac{m\pi x}{W}
                \end{align}
            \end{subequations}
            where we define $A_m=A_{m,0}$.  This approximation combined with the conditions Eqs.\ (\ref{eqapp:sm}) and (\ref{eqapp:Amn even}) for $n=0$ allows us to write in closed-form
            \begin{equation}
                A_m=-\frac{2s_m\sinh\left(\frac{m\pi H}{2W}\sqrt{\frac{Y_x}{Y_y}}\right)}{H\left(\frac{m^2\pi^2B_x}{L_x^2Y_y}+1\right)}
            \end{equation}
            and
            \begin{widetext}
                \begin{equation}
                    s_m=\frac{2B_x\kappa_xH\sin\frac{m\pi}{2}}{WY_y\sinh\left(\frac{m\pi H}{2W}\sqrt{\frac{Y_x}{Y_y}}\right)\left[\frac{m\pi H}{2W}\sqrt{\frac{Y_x}{Y_y}}\left(\frac{m^2\pi^2B_x}{W^2Y_y}+1\right)\cosh\left(\frac{m\pi H}{2W}\sqrt{\frac{Y_x}{Y_y}}\right)-1\right]}
                \end{equation}
            \end{widetext}
            Substituting this approximation in the elastic energy functional, we obtain
            \begin{equation}
                \tilde{\mathcal{E}}_{\rm ex}(w,h)=1-\frac{2}{w^2}\sum_{m\textrm{ odd}}\frac{1}{\frac{m^2\pi^2}{4w^2}+1-\frac{2w}{m\pi h}\tanh\frac{m\pi h}{2w}}
                \label{eqapp:excess energy tall}
            \end{equation}
            where we rescaled by the flattening energy $\mathcal{E}_{\rm flat}=B_x\kappa_x^2/2$.

            We can check several limits.  For tall assemblies with $h\rightarrow\infty$, we find
            \begin{equation}
                \tilde{\mathcal{E}}_{\rm ex}^{(\rm tall)}(w)=1-\frac{2}{w^2}\sum_{m\textrm{ odd}}\frac{1}{\frac{m^2\pi^2}{4w^2}+1}=1-\frac{\tanh w}{w}
            \end{equation}
            For wide assemblies with $w\rightarrow\infty$ (equivalently, an annulus), we find
            \begin{equation}
                \tilde{\mathcal{E}}_{\rm ex}^{(\rm wide)}(h)=1-\frac{24}{\pi^2(3+h^2)}\sum_{m\textrm{ odd}}\frac{1}{m^2}=\frac{h^2}{3+h^2}
            \end{equation}
            Finally for small assemblies with $w,h\ll1$, we have to leading order
            \begin{equation}
                \tilde{\mathcal{E}}_{\rm ex}(w,h)\simeq\left(\frac{1}{3}-\frac{64w}{\pi^5h}\sum_{m\textrm{ odd}}\frac{1}{m^5}\tanh\frac{m\pi h}{2w}\right)w^2
            \end{equation}
            We briefly prove that this is symmetric about $h=w$.  We use the following identity (see Eq.\ (\ref{eqapp:excess energy tall}))
            \begin{equation}
                \frac{\tanh z}{z}=\sum_{m\textrm{ odd}}\frac{2}{\frac{m^2\pi^2}{4}+z^2}
            \end{equation}
            The excess energy can then be rewritten as
            \begin{align}
                \begin{split}
                    \tilde{\mathcal{E}}_{\rm ex}(w,h)&\simeq\left(1-\frac{96}{\pi^4}\sum_{m\textrm{ odd}}\frac{1}{m^4}\frac{\tanh\frac{m\pi h}{2w}}{\frac{m\pi h}{2w}}\right)\frac{w^2}{3}\\
                    &=\left(1-\frac{768}{\pi^6}\sum_{m,n\textrm{ odd}}\frac{1}{m^4}\frac{w^2}{n^2w^2+m^2h^2}\right)\frac{w^2}{3}
                \end{split}
            \end{align}
            Using
            \begin{equation}
                \frac{w^2}{n^2w^2+m^2h^2}=\frac{1}{n^2}-\frac{m^2h^2}{n^2(n^2w^2+m^2h^2)}
            \end{equation}
            and
            \begin{equation}
                \sum_{m\textrm{ odd}}\frac{1}{m^4}=\frac{\pi^4}{96}; \sum_{n\textrm{ odd}}\frac{1}{n^2}=\frac{\pi^2}{8}
            \end{equation}
            we finally write the excess energy in the small size limit as
            \begin{equation*}
                \tilde{\mathcal{E}}_{\rm ex}(w\ll1,h\ll1)\simeq\frac{1}{3}\sum_{m,n\textrm{ odd}}\frac{1}{m^2n^2\left(\frac{m^2}{w^2}+\frac{n^2}{h^2}\right)}
            \end{equation*}
            This is symmetric about $h=w$.
            
	\section{\label{app:mesoscopic curvature}Mesoscopic curvature of wide ribbons}
		We here derive a discrete version of the centerline curvatures of wide or annular assemblies.  Consider an annular assembly that is $N$ particles thick.  Since this annular assembly is radially symmetric, we can treat it as a ring of radial stacks of particles.  Suppose that the inner ring has radius $R$ and adjacent radial stacks are simply rotated relative to each other by an angle $\delta\theta$ around the center of the annulus.  For simplicity (and from observation of simulations of wide assemblies), we assume that there is not significant radial compression of the rings i.e.\ $u_r\ll R$.  The elastic energy between two adjacent rows is simply
        \begin{widetext}
            \begin{multline}
                E=\sum_{n=1}^{N}\frac{1}{2}k_xa^2\left[2\left(\frac{R}{a}+n-1\right)\sin\frac{\delta\theta}{2}-(1+\tan\alpha)\cos\frac{\delta\theta}{2}+\sin\frac{\delta\theta}{2}\right]^2\\
                +\sum_{n=1}^{N}\frac{1}{2}k_xa^2\left[2\left(\frac{R}{a}+n-1\right)\sin\frac{\delta\theta}{2}-(1-\tan\alpha)\cos\frac{\delta\theta}{2}-\sin\frac{\delta\theta}{2}\right]^2.
            \end{multline}
        \end{widetext}
        Minimizing this energy with respect to $R$ and $\delta\theta$, we have
        \begin{equation}
            R=\frac{a}{2\tan\frac{\delta\theta}{2}}-\frac{(N-1)a}{2}=R_c-\frac{(N-1)a}{2},
        \end{equation}
        and
        \begin{multline}
            \tan\frac{\delta\theta}{2}=\frac{1}{2}\left[-\frac{N^2+2-3\tan^2\alpha}{3\tan\alpha}\right.\\
            \left.+\sqrt{\left(\frac{N^2+2-3\tan\alpha}{3\tan\alpha}\right)^2}+4\right]
        \end{multline}
        where $R_c$ is the centerline radius.  Note that for $N=1$, we have that the centerline radius is $R_c=a/(2\tan\alpha)=\kappa_0^{-1}$, which is exactly the preferred radius of a single ring of particles.  For $N\gg1$, we have $R_c\simeq N^2a/(6\tan\alpha)=N^2/(3\kappa_0)$.  So the centerline curvature scales goes as $\kappa_c\simeq\kappa_0/(N^2/3)$.
		
		This scaling of the centerline curvature with thickness can also be obtained from a simple energy argument.  The bending energy for an annulus with curvature $\kappa$ is $\mathcal{E}_{\rm bend}\sim B(\kappa-\kappa_0)^2$.  The amount of extension or compression energy near the outer and inner rings can be estimated as $\mathcal{E}_{\rm stretch}\sim Y(\kappa H)^2$.  Minimizing the sum of these energies with respect to $\kappa$, we find
        \begin{equation}
            \kappa_c=\frac{\kappa_0}{1+\gamma h^2}
        \end{equation}
        where $h=H/\lambda_y=H/(2\sqrt{B_x/Y_x})$ and $\gamma$ is some constant.  Comparing with the discrete limit, we find $\gamma=1/3$.

	\bibliography{refs.bib}
	
\end{document}